\documentclass[11pt]{article}
\usepackage[utf8]{inputenc}
\usepackage[a4paper, total={6in, 8in}]{geometry}
\usepackage{mathrsfs}
\usepackage{amssymb}
\usepackage{braket}
\usepackage{amsmath}
\usepackage{graphicx}
\usepackage{comment}

\usepackage[
backend=bibtex,
style=phys,
biblabel=brackets,
abbreviate=false,
doi=false,
url=false,
isbn=false,
eprint=true,
]{biblatex}
\AtEveryBibitem{%
	\clearfield{note}%
}

\usepackage[T1]{fontenc}
\usepackage{lmodern}
\usepackage{morefloats}
\usepackage[svgnames]{xcolor}
\usepackage{tikz}
\usetikzlibrary{calc,decorations.markings, shapes.misc, decorations.pathmorphing}
\tikzset{cross/.style={cross out, draw, 
		minimum size=2*(#1-\pgflinewidth), 
		inner sep=0pt, outer sep=0pt}} 
\usepackage{tqft}
\definecolor{lstbgcolor}{rgb}{0.9,0.9,0.9} 
\usepackage{listings}
\def\cO{ \mathcal{O}} 
\usepackage{fancyvrb}

\newcommand{\normord}[1]{%
	{{#1}}%
}

\newcommand{\bea}{\begin{eqnarray}}
	\newcommand{\eea}{\end{eqnarray}}
\newcommand{\bg}{\begin{gathered}}
	\newcommand{\eg}{\end{gathered}}

\numberwithin{equation}{section}
\allowdisplaybreaks
\usepackage{subcaption}

\usepackage{amsthm}
\usepackage{rotating}
\usepackage{physics}
\usepackage{tensor}
\makeatletter
\newcommand{\bigtimes}{%
	\DOTSB\mathop{\mathpalette\mattos@bigtimes\relax}\slimits@
}
\newcommand\mattos@bigtimes[2]{%
	\vcenter{\hbox{%
			\sbox\z@{$#1\sum$}%
			\resizebox{!}{0.9\dimexpr\ht\z@+\dp\z@}{\raisebox{\depth}{$\m@th#1\times$}}%
	}}%
	\vphantom{\sum}%
}
\makeatother

\newcommand{\merge}{\vee}
\newcommand{\invmap}{Q}

\DeclareMathOperator{\con}{}

\DeclareMathOperator{\Dim}{Dim}
\DeclareMathOperator{\Mult}{Mult}

\newcommand{\Sym}{\mathrm{Sym}}

\definecolor{GREEN}{rgb}{0.0,0.70,0.24}
\definecolor{BLUE}{rgb}{0.0,0.24,0.70}

\newcommand{\PAdiagramLabeled}[2]{\;\begin{tikzpicture}[baseline={([yshift=-.5ex]current bounding box.center)}]
		\def \n {#1};
		\def \edges {#2};
		\def \sep {0.5};
		\foreach \v in {1,...,\n}
		{
			\pgfmathparse{(\v-1)*\sep};
			\coordinate (v\v) at (\pgfmathresult,0.25);
			\node[circle,fill,inner sep=1pt, label=above:{\tiny $\v'$}] at (v\v) {};
		}
		\foreach \v in {1,...,\n}
		{
			\pgfmathparse{(\v-1)*\sep};
			\coordinate (v-\v) at (\pgfmathresult,-0.25);
			\node[circle,fill,inner sep=1pt, label=below:{\tiny $\v$}] at (v-\v) {};
		}
		\foreach \endpointOne/\endpointTwo in \edges
		{
			\draw[] (v\endpointOne) -- (v\endpointTwo);
		}
	\end{tikzpicture}\;}

\newcommand{\PAdiagram}[3][]{\;\begin{tikzpicture}[baseline={([yshift=-.5ex]current bounding box.center)}]
		\def \n {#2};
		\def \edges {#3};
		\def \arcs {#1}
		\def \sep {0.5};
		\foreach \v in {1,...,\n}
		{
			\pgfmathparse{(\v-1)*\sep};
			\coordinate (v\v) at (\pgfmathresult,0.25);
			\node[circle,fill,inner sep=1pt] at (v\v) {};
		}
		\foreach \v in {1,...,\n}
		{
			\pgfmathparse{(\v-1)*\sep};
			\coordinate (v-\v) at (\pgfmathresult,-0.25);
			\node[circle,fill,inner sep=1pt] at (v-\v) {};
		}
		\foreach \endpointOne/\endpointTwo in \edges
		{
			\draw[] (v\endpointOne) -- (v\endpointTwo);
		}
		\foreach \endpointOne/\endpointTwo in \arcs
		{
			\draw[] (v\endpointOne) to[bend left] (v\endpointTwo);
		}
	\end{tikzpicture}\;}

\def\ls[#1]{ {}_{#1}}
\def\us[#1]{ \mbox{\tiny{#1}}}

\usepackage{hyperref}
\hypersetup{
	colorlinks,
	linkcolor={red!50!black},
	citecolor={blue!50!black},
	urlcolor={blue!80!black}
}

\bibliography{bibliography}

\title{\Large \bf Hidden symmetries and Large $N$ factorisation for permutation invariant matrix observables}
\date{}
\author{{George Barnes$^{a,}$\footnote{g.barnes@qmul.ac.uk}, Adrian Padellaro$^{a,}$\footnote{a.k.s.padellaro@qmul.ac.uk}, Sanjaye Ramgoolam$^{a , b,}$\footnote{s.ramgoolam@qmul.ac.uk}  }}
\begin{document}
\begin{flushright}
	QMUL-PH-21-46
\end{flushright}
{\let\newpage\relax\maketitle}
\vspace{-2.5em}
\begin{center}
	$^{a}${\em  Centre for Theoretical Physics}, {\em Department  of Physics and Astronomy}, \\
	{\em Queen Mary University of London}, \\
	{\em London E1 4NS, United Kingdom }\\
	
	\medskip
	$^{b}${\em  School of Physics and Mandelstam Institute for Theoretical Physics,} \\   
	{\em University of Witwatersrand}, \\ 
	{\em Wits, 2050, South Africa} \\
	\medskip	
\end{center}
\begin{abstract}
	Permutation invariant polynomial functions of matrices have previously been studied  as the observables in matrix models invariant under $S_N$, the symmetric group of all permutations of $N$ objects. In this paper, the permutation invariant matrix observables (PIMOs) of degree $k$ are shown to be in one-to-one correspondence with equivalence classes of elements in the diagrammatic  partition algebra $P_k ( N )$. On a 4-dimensional subspace of the 13-parameter space of $S_N$ invariant Gaussian  models, there is an  enhanced $O(N)$ symmetry. At a special point in this subspace, is the simplest $O(N)$ invariant action. This is used to define an inner product on the PIMOs which is expressible as a trace of a product of elements in the partition algebra.  The diagram algebra  $P_k ( N )$ is used to prove the large $N$ factorisation  property of this inner product, which generalizes a familiar large $N$ factorisation  for inner products of matrix traces invariant under continuous symmetries. 
\end{abstract}
\newpage
\tableofcontents
\section{Introduction}

The simplifications of large $N$ in  matrix quantum field theories in diverse dimensions with continuous gauge symmetries such as $U(N), O(N), Sp(N)$ discovered in \cite{tHooft} have played a major role in the development of  gauge-string duality in subsequent years. This includes  low-dimensional non-critical string theories dual to zero-dimensional QFTs (matrix models) \cite{Douglas:1989ve,Brezin1990,GrossMigdal}, the string dual of two-dimensional Yang-Mills theories \cite{GrTa}, and the generalization to higher dimensions in the AdS/CFT correspondence \cite{Malda}.  
In theories with continuous symmetries containing adjoint fields, the space of
gauge invariants is generated by traces of matrices. An important aspect of  simplicity in 
the large $N$ limit  is ``large $N$ factorisation''. In the context of AdS/CFT, large $N$ factorisation for two-point functions  involving gauge invariants built from a complex matrix  is an expression of orthogonality for distinct trace structures \cite{BBNS}. This plays an important role in the connection between multi-traces constructed from a small number of matrices and perturbative gravitons in the AdS dual \cite{Witten1998,GKP}. The breakdown of this orthogonality when the number of matrices becomes comparable to $N$ guides the  identification of CFT duals \cite{BBNS,CJR,Berenstein2004}  for giant gravitons \cite{mst,HHI2000,GMT2000}. Large $N$ factorisation also enters the  construction of gauge-string duals in collective field theory \cite{JevSak,Yaffe1982,DasJev,RobJev}, which gives useful insights into the emergence of classical limits at large $N$. It is  also employed in the Master field approach to large $N$ \cite{Witten1980} and loop equations \cite{MigMak} (see for example \cite{Lin2020,KZ2108,KJLM} for recent developments of these themes).   In the geometrical construction of gauge-string duality, based on Schur-Weyl duality and branched covers, in instances such as large $N$ $2d$ Yang Mills \cite{1993Gross_1, Minahan1993,  GrTa, SCHNITZER1993, Gross1993_3, MP1993, Horava1996, CMR, Kimura:2008gs} or the simple toy model of Gaussian Hermitian matrix theory \cite{ITZYK, RobSanj, Gopak2011, dMKLN}, trace structures of matrix invariants correspond to branching structures of covering maps from string worldsheets. 

In this paper, we will develop the theme of large $N$ factorisation for permutation invariant matrix models \cite{LMT, PIGMM, Ramgoolam2019, PIG2MM}.    Permutation invariance was motivated in \cite{LMT} in the context of matrix data arising in computational linguistics \cite{coecke2010mathematical, maillard2014type, Baroni2014FregeIS, Grefenstette2015ConcreteMA}. 
The formulation of large $N$ factorisation we will use is similar to the one in \cite{BBNS}. We will use the simplest inner product on the space of permutation invariant matrix observables (PIMOs). It comes from a special point on the moduli space of $S_N$ invariant Gaussian matrix models where the action has $O(N)$ symmetry.  This is the first sense in which hidden symmetries appear in this paper. Permutation invariant random matrix distributions have also been studied from the point of view of mathematical statistics, using partition algebra diagrams \cite{gabrielcombinatorial1, gabrielcombinatorial2, gabrielcombinatorial3}.

The second kind of hidden symmetry appearing in this paper is based on 
 Schur-Weyl duality. Observables invariant under the  action of a symmetry group $G$, as we will discuss in this paper,  are organized by  algebras dual to $G$. For the case of $U(N)$ symmetry the dual algebras are based on the standard Schur-Weyl  duality \cite{FultonHarris} between 
 $U(N)$ and $S_k $ in the $k$-fold tensor product of  $V^{ \otimes k } $ of the fundamental representation $V$ of $U(N)$. Applications of Schur-Weyl duality to the computation of correlators in matrix models with $U(N)$ symmetry are developed in \cite{CJR,Kimura2007a, Brown2008, Bhattacharyya2008, Bhattacharyya2008b, Brown2009, Kimura2008,QuivCalc,Ber1504,CDD1301,KRS,LY2107,ADHSSS,CLBSR}    and short reviews are \cite{SRrev1,Ramgoolam2016}.  The $U(N)$ case  serves as a powerful source of useful  analogies throughout the paper.
 When $U(N)$ is replaced by $S_N$  as the invariance of interest, the Schur-Weyl dual algebras are diagrammatic partition algebras $P_k(N)$.  These algebras have been studied in statistical physics and representation theory  
 \cite{Martin1994, Martin1996,Halverson2004}. The algebras $P_k(N)$ are finite dimensional and have a  basis which can be labelled by diagrams, corresponding to  set partitions of a set of $2k$ objects.  A set partition of a set $S$ is a collection of non-empty subsets of $S$,  such that any pair of of subsets has zero intersection, and the union of the subsets is the set $S$. Equivalently, every  element of the set is included in exactly one of the subsets (see for example \cite{enwiki:1054411091}  for further information on set partitions).

The paper is organised as follows. In section \ref{sec: PIGMM} we review the construction of the permutation invariant Gaussian 1-matrix model, and the counting of invariant matrix observables developed in \cite{LMT, PIGMM}. Here we give a new description of the counting, which emphasizes the underlying hidden partition algebra symmetry arising as a consequence of Schur-Weyl duality.  We end the section with a derivation of the $O(N)$ symmetric point in the moduli space of $S_N$ invariant 1-matrix models.

Section \ref{sec: PIMOs} is dedicated to the construction of PIMOs by means of partition algebras. We give a brief description of the partition algebras. In particular we present the diagram basis and describe how the product is computed by using a composition of diagrams. The construction of $U(N)$ invariants using symmetric group algebras is reviewed as a warm-up exercise.
This is generalised to give a map from partition algebra elements to PIMOs (equation \eqref{eq: PIMO from PkN}), leading to a correspondence between PIMOs and equivalence classes of partition algebra elements. These equivalence classes are defined  in equation \eqref{Equivs}.  
The simplest $O(N)$ invariant action is used to define an  inner product 
on the space of PIMOs in terms of a trace of partition algebra elements (equation \eqref{eq: Two Point Function of Partition Algebra Observables}).

Section \ref{sec: factorization} proves the  large $N$ factorization of the  inner product  on PIMOs thus defined. That is, we show that 
\begin{equation}
	\expval{\hat{O}_i\hat{O}_j} = \delta_{ij} + O(1/\sqrt{N}),
\end{equation}
where $\hat{O}_i, \hat{O}_j $ are normalized PIMOs labelled by indices $i,j$ running over  equivalence classes of partition algebra elements.
The proof of large $N$ factorization relies on the existence of a partial ordering on the diagram basis for the partition algebra. The partial ordering is  related to an inclusion  of diagrams, and can itself be described by another diagram of diagrams called a Hasse diagram \cite{birkhoff1940lattice}. We end the section by extending the proof  to multi-matrix observables.

\section{Hidden symmetries in permutation invariant Gaussian matrix models}
\label{sec: PIGMM}
In \cite{PIGMM} a 13-parameter family of Gaussian matrix models consistent with permutation invariance was constructed, by using a transformation from the matrix variables $M_{ij}$ to variables labelled by irreducible representations of $S_D$. The expectation values of linear and quadratic permutation invariant polynomials in $M_{ij}$ were given in terms of the representation theoretic parameters. Expectation values for a sample of cubic and quartic invariant polynomials  were constructed using Wick's theorem. Additional examples were computed in \cite{Ramgoolam2019}. The results were generalized to the 2-matrix case in 
\cite{PIG2MM}.  Computer code for expectation values of invariant polynomials in the 1-matrix and 2-matrix case is available as part of \cite{PIG2MM}. 

The schematic form of the permutation invariant Gaussian matrix model (PIGMM) is
\begin{equation}\label{action} 
	\int \dd{M} \exp({-S(M)}) = \int \dd{M} \exp(-\sum_{i=1}^2\mu_i L_i(M) - \sum_{i=1}^{11} g_{i} Q_i(M)).
\end{equation}
The action $S(M)$ contains two linear terms: $L_1, L_2$; and eleven quadratic terms $Q_1, \dots Q_{11}$.
It is the most general quadratic action invariant under the following group action of $S_N$ (the symmetric group on $N$ objects),
\begin{equation}
	S(M_{\sigma(i)\sigma(j)}) = S(M_{ij}), \qq{for $ \hbox{ all } \sigma \in S_N$.} \label{eq: PIGMM invariance}
\end{equation}

The permutations  $\sigma \in S_N$ are invertible maps $\sigma: \{1,\cdots,N\} \rightarrow \{1,\dots,N\}$. The product of two  permutations $ \sigma_1 , \sigma_2$ is defined by composing the maps : $\sigma_1 \sigma_2 (i) =  \sigma_2(\sigma_1(i))$.  As an  example, consider the following two permutations in $S_3$: 
\begin{equation}
	\begin{aligned}
		&\sigma_1: 1 \mapsto 2, \, 2 \mapsto 3, \, 3 \mapsto 1, \\
		&\sigma_2: 1 \mapsto 2, \, 2 \mapsto 1, \, 3 \mapsto 3.
	\end{aligned}
\end{equation}
In this case 
\begin{equation}
	\sigma_1 \sigma_2: 1 \mapsto 1, \, 2 \mapsto 3, \, 3 \mapsto 2.
\end{equation}
Let $V_N$ be an $N$-dimensional vector space with orthonormal basis $e_i$, $i=1,\dots,N$.
The defining representation $\rho_N: S_N \rightarrow GL(V_N)$ of $S_N$ assigns the following linear operator to every permutation
\begin{equation}
	\rho_N(\sigma)e_i = e_{\sigma^{-1}(i)}.
\end{equation}

From equation \eqref{eq: PIGMM invariance}, we see that the vector space spanned by $M_{ij}$ is acted on by $S_N$ in the same way as $V_N \otimes V_N$. We have the identification
\begin{equation}
	M_{ij} \longleftrightarrow e_i \otimes e_j.
\end{equation}
This is not an irreducible representation, it decomposes into several irreducible components
\begin{equation}
	V_N \otimes V_N \cong 2 V_{[N]} \oplus 3 V_{[N-1,1]} \oplus V_{[N-2,1,1]} \oplus V_{[N-2,2]}. \label{eq: V_N times V_N Decomposition}
\end{equation} 
Here $V_{[N]}$ corresponds to the trivial one-dimensional  representation of $S_N$. The representations   $V_{[N-1,1]}, V_{[N-2,1,1]}, V_{[N-2,2]}$ are non-trivial irreducible representations, labeled by integer partitions of $N$.
Their dimensions are respectively 
\bea 
N -1 , ( N -1 ) ( N -2  ) /2 , N ( N -3 ) /2 . 
\eea
We will use the index $\Lambda_1$ ranging over the labels for irreducible representations 
\bea 
	\Lambda_1 \in \qty {  {[ N]},{[N-1,1]}, {[N-2,1,1]}, {[N-2,2]}  } \, 
\eea
and we refer to the corresponding irreducible representations of $S_N$ as  $V^{S_N}_{\Lambda_1}$.
The above decomposition can be deduced using
\begin{equation}
	V_N \cong V_{[N]} \oplus V_{[N-1,1]},
\end{equation}
together with the tensor product rule described in section 7.13 of \cite{Hamermesh1962}. See also \cite{Sagan2013} for a dedicated treatment of  symmetric group representation theory. 

Note that the multiplicity of $V_{[N]}$ in \eqref{eq: V_N times V_N Decomposition} is exactly why there are two linear terms $L_1, L_2$ in the action \eqref{action}. Furthermore
the isomorphism in equation \eqref{eq: V_N times V_N Decomposition} implies that there exists a set of linear combinations of matrix elements labelled by $ \Lambda_1$ 
\begin{equation}
	S^{\Lambda_1, \alpha}_a = \sum_{i,j}C^{\Lambda_1, \alpha}_{a,ij}M_{ij}. \label{eq: Irrep Basis Matrix Elements}
\end{equation}
The index $a$ is a state index for the irreps, while $\alpha$ is a multiplicity index
\begin{equation}
	\begin{aligned}	
		&a \in \{  1,\dots, \Dim V_{\Lambda_1}^{S_N} \} , \\
		&\alpha \in \{  1,\dots, \Mult(V_N \otimes V_N \rightarrow V_{\Lambda_1}^{S_N}) \} ,
	\end{aligned}
\end{equation}
where $\Mult(V_N \otimes V_N \rightarrow V_{\Lambda_1}^{S_N})$ is the multiplicity of $V_{\Lambda_1}^{S_N}$ in $V_N \otimes V_N$.
The change of basis is given by the Clebsch-Gordan coefficients $C^{\Lambda_1, \alpha}_{a,ij}$.
They have the property
\begin{equation}
	\sum_{i,j}C^{\Lambda_1, \alpha}_{a,ij}M_{\sigma^{-1}(i)\sigma^{-1}(j)} = \sum_b D^{\Lambda_1}_{ab}(\sigma)S^{\Lambda_1 \alpha}_b,
\end{equation}
where the matrices $D^{\Lambda_1}_{ab}(\sigma)$ are irreducible matrix representations of $S_N$ 
(background on the Clebsch-Gordan coefficients for symmetric groups is available in  \cite{Hamermesh1962}). 
Without loss of generality, we can assume that the Clebsch-Gordan coefficients define an orthonormal basis with respect to the inner product
\begin{equation}
	(M_{ij}, M_{kl}) = \delta_{ik}\delta_{jl}. \label{eq: inner product V_N otimes V_N}
\end{equation}
Equivalently, the representation theoretic variables  satisfy
\begin{equation}
	(S_a^{\Lambda_1, \alpha}, S_{b}^{\Lambda_1',\beta}) = \delta_{ab}\delta^{\Lambda_1 \Lambda_1'}\delta^{\alpha \beta}.
\end{equation}
Together with the fact that the inner product \eqref{eq: inner product V_N otimes V_N} is $S_N$ invariant, it follows that
\begin{equation}
	D^{\Lambda_1}_{ab}(\sigma^{-1}) = D^{\Lambda_1}_{ba}(\sigma).
\end{equation}

Using the above basis it immediately follows that the quadratic combination
\begin{equation}
	\sum_{a} S^{\Lambda_1, \alpha}_a S^{\Lambda_1, \beta}_a = \sum_{i,j,k,l} M_{ij}\invmap^{\Lambda_1, \alpha \beta}_{ijkl}M_{kl}
\end{equation}
is an invariant polynomial, where
\begin{equation}
	\invmap^{\Lambda_1, \alpha \beta}_{ijkl} = \sum_{a} C_{a, ij}^{{\Lambda_1}, \alpha} C_{a, kl}^{{\Lambda_1}, \beta}.
\end{equation}
A useful observation is that, while the Clebsch-Gordan coefficients depend on a choice of basis for every irreducible component in \eqref{eq: V_N times V_N Decomposition}, the tensors $\invmap^{\Lambda_1, \alpha \beta}_{ijkl}$ do not. For all four $\Lambda_1$, they can be constructed by using only  the explicit bases for the subspaces $V_{ [N] } $ and $ V_{[N-1,1]}$  in \eqref{eq: V_N times V_N Decomposition} \cite{PIGMM}.

We may associate a unique parameter to every invariant. Since
\begin{equation}
	\sum_{i,j,k,l} M_{ij}\invmap^{\Lambda_1, \alpha \beta}_{ijkl}M_{kl} = \sum_{i,j,k,l} M_{ij}\invmap^{\Lambda_1, \beta \alpha}_{ijkl}M_{kl},
\end{equation}
there is a symmetric matrix of dimension $\Mult(V_N \otimes V_N \rightarrow V_{\Lambda_1}^{S_N})$ parametrising the quadratic part of the action, for every choice of $\Lambda_1$. Using the multiplicities in the decomposition \eqref{eq: V_N times V_N Decomposition}, we have
\begin{equation}
	11 = \frac{2\cdot 3}{2!} + \frac{3 \cdot 4}{2!} + \frac{1\cdot 2}{2!} +  \frac{1\cdot 2}{2!}.
\end{equation}
independent parameters.
The two linear terms are given by
\begin{equation}
	\mu_1  L_1 = \mu_1  S^{[N],1} \qq{and}  \mu_2  L_2 =  \mu_2 S^{[N],2}.
\end{equation}
The quadratic part is
\begin{equation}
	\sum_{\substack{\Lambda_1,a\\ \alpha, \beta}}S^{\Lambda_1, \alpha}_a g^{\Lambda_1}_{\alpha \beta}S^{\Lambda_1, \beta}_a = \sum_{\substack{i,j,k,l \\ \Lambda_1, \alpha, \beta}} g^{\Lambda_1}_{\alpha \beta} M_{ij}\invmap^{\Lambda_1, \alpha \beta}_{ijkl}M_{kl}.
\end{equation}
where the matrices  $g^{ \Lambda_1}_{\alpha \beta } $ are parameters of the model. 
In this basis the partition function is,
\begin{equation} \label{eq: SN partition fn}
	\begin{aligned}
		\int \dd{M} \exp(-S(M)) = \int &\dd{S} \exp(-\mu_1 S^{[N],1} - \mu_2 S^{[N],2}-\sum_{\substack{\Lambda_1,a\\ \alpha, \beta}}S^{\Lambda_1, \alpha}_a g^{\Lambda_1}_{\alpha \beta}S^{\Lambda_1, \beta}_a) \\
		&\dd{S} = \prod_{\Lambda_1, \alpha, a} \dd{S_a^{\Lambda_1, \alpha}}.
	\end{aligned}
\end{equation}
The matrices $g^{\Lambda_1}_{\alpha \beta}$ must have non-negative eigenvalues to define a convergent integral.

Note that the parameters in the quadratic part of the action in \cite{PIGMM} are related to the parameters in this paper as
\begin{equation}\label{eq: PIGMM irrep labels}
	\begin{aligned}
		&\big( \Lambda_{V_0} \big)_{\alpha \beta}  \longleftrightarrow g^{[N]}_{\alpha \beta},\\
		&\big( \Lambda_{V_H} \big)_{\alpha \beta}  \longleftrightarrow g^{[N-1,1]}_{\alpha \beta}, \\
		&\big( \Lambda_{V_2} \big) \longleftrightarrow g^{[N-2,2]}, \\
		&\big( \Lambda_{V_3} \big)  \longleftrightarrow g^{[N-2,1,1]}.
	\end{aligned}
\end{equation}
 The  slight shift of notation makes the connection between the parameters and the decomposition \eqref{eq: V_N times V_N Decomposition} more manifest.

\subsection{Counting matrix observables using partition algebras}

 Permutation invariant matrix polynomials, which have been studied as the natural class of observables in the matrix model with permutation invariant measure and action, are defined to obey  
\begin{equation}
	\mathcal{O}(M_{\sigma(i)\sigma(j)}) = \mathcal{O}(M_{ij}) \hbox{ for all } \sigma \in S_N 
\end{equation}
These permutation invariant matrix observables (PIMOs) can be  organized by their degree.
At degree $k$, the matrix monomials
\begin{equation}
	M_{i_1 i_{1'}} M_{i_2 i_{2'}} \dots M_{i_k i_{k'}},
\end{equation}
form a basis for a vector space isomorphic to $\Sym^k(V_N \otimes V_N)$. The symmetric group $S_k$ acts on $(V_N \otimes V_N)^{\otimes k}$ by permuting the $k$ tensor factors. The subspace $\Sym^k(V_N \otimes V_N)$ is the subspace of $S_k$ invariants in $(V_N \otimes V_N)^{\otimes k}$. This $S_k$ invariance is  imposed because of the bosonic symmetry of the matrix variables  $M_{ij}$. The PIMOs form the $S_N \times S_k$ invariant subspace of $(V_N \otimes V_N)^{\otimes k}$ : 
\bea
&& \hbox{ Matrix polynomials of degree $k$ invariant under $S_N$  }\cr 
&& = \hbox{Invariants}_{~ S_{N } \times S_k }  \left ( (V_N \otimes V_N)^{\otimes k} \right ) \equiv [(V_N \otimes V_N)^{\otimes k}]^{S_N \times S_k}
 \cr 
&&  = \{ v \in (V_N \otimes V_N)^{\otimes k} : \sigma v = v, \tau v = v \, \vert \, \hbox{ for all }   \sigma \in S_N, \tau \in S_k\}.
\eea
Note that the action of $\tau \in S_k$ on $(V_N \otimes V_N)^{\otimes k}$ commutes with the action of $\sigma \in S_N$. This follows since the same $\sigma$ is applied to all tensor factors.

In \cite{LMT} the dimension of the space  of independent PIMOs for matrices of size $N$  and polynomial  degree $k$ was obtained as 
\begin{equation}
	\mathcal{N}(N, k) =  \frac{1}{N! k!} \sum_{p \vdash N} \sum_{q \vdash k}  \frac{N!}{\prod_{i=1}^N i^{p_i} p_i!} \frac{k!}{\prod_{i=1}^k i^{q_i} q_i!} \prod^{k}_{i=1} \big(\sum_{l|i} l p_l \big)^{2q_i} \label{eq: Counting 1-matrix Invariants using Projectors}.
\end{equation}
 The initial sums run over integer partitions (Young diagrams) $p$ of $N$, and integer partitions $q$ of $k$ while the final sum is over the integer divisors $l$ of $i$.
The  equation \eqref{eq: Counting 1-matrix Invariants using Projectors} computes the multiplicity of the trivial representation of $S_N \times S_k$ in the decomposition of $(V_N \otimes V_N)^{\otimes k}$, which is the dimension of $[(V_N \otimes V_N)^{\otimes k}]^{S_N \times S_k}$.
 There exists an isomorphism
\begin{equation}
	(V_N \otimes V_N)^{\otimes k} \cong \bigoplus_{\Lambda_1, \Lambda_2} V_{\Lambda_1}^{S_N} \otimes V_{\Lambda_2}^{S_k} \otimes V_{\Lambda_1 \Lambda_2},
\end{equation}
organizing the space into irreducible representations of $S_N \times S_k$, with multiplicities $V_{\Lambda_1 \Lambda_2}$. Let $V_{[N]} \otimes V_{[k]}$ denote the trivial representation of $S_N \times S_k$ with multiplicity space  $V_{[N] [k]}$, then the dimension of $S_N \times S_k$ invariants is given by 
\begin{equation}
	\mathcal{N}(N,k) = \Dim V_{[N] [k]}.
\end{equation}
The generalization to multi-matrix observables and a proof of their correspondence with colored directed graphs was developed in \cite{PIG2MM}. The approach in this paper is based on a new way of counting PIMOs, utilising the connection between dual algebras and matrix invariants.

We begin by reviewing this connection in the case of $U(N)$ invariants. 
 Tensor products of the defining representation $V$ of $U(N)$ have a multiplicity free  decomposition into irreducible representations of $ U(N) \times S_k$ labelled by Young diagrams 
\begin{equation} \label{eq: U(N) S_k duality}
	V^{\otimes k} \cong \bigoplus_{\substack{\Lambda \vdash k \\ l(\Lambda) \leq N}} V^{U(N)}_{\Lambda} \otimes V^{\mathbb{C}S_k}_{\Lambda}.
\end{equation}
The sum runs over Young diagrams $\Lambda $  with $k$ boxes, and for $k>N $ is restricted such that the number of rows  $ l ( \Lambda)$   in the Young diagram $\Lambda $  is not  greater than $N$.
In the remainder of this paper we will assume $N \geq k$ for discussions of the unitary group.
This result is known as Schur-Weyl duality (see chapter 6 in \cite{FultonHarris}). On the left-hand side of this equation we have a basis $e_{i_1} \otimes e_{i_2} \otimes \dots \otimes e_{i_k}$ with each index $i$ running from $1$ to $N$. On the right-hand side we have a basis $E^{\Lambda}_{M m}$ with 
\begin{equation}
	\begin{aligned}
		&m \in \{ 1, \dots, \Dim V_{\Lambda}^{\mathbb{C}S_k} \} ,\\
		&M \in \{ 1,  \dots, \Dim V_{\Lambda}^{U(N)}\} . 
	\end{aligned}
\end{equation}
For a  fixed Young diagram  $ \Lambda$ and a fixed state  $ M $ in $V_{\Lambda}^{U(N)}$, 
there is a multiplicity of $\Dim V_{\Lambda}^{\mathbb{C}S_k} $.
That is, we have
\begin{equation}
	\Mult(V^{\otimes k} \rightarrow V^{U(N)}_{\Lambda}) = \Dim V_{\Lambda}^{\mathbb{C}S_k}.
\end{equation}

It is well-known that $U(N)$ invariant matrix observables have a basis of multi-traces.
These traces can be parameterised by conjugacy classes of permutations. A description of 
the connection between gauge invariant observables and equivalence classes of permutations
for single matrix as well as multi-matrix problems, with applications to AdS/CFT is given in \cite{Ramgoolam2016}. We review the connection here with an emphasis on  Schur-Weyl duality from the outset.  This framework, as explained in \cite{Ramgoolam2016}, can be used to give a description of finite $N$ effects on the counting and construction of gauge invariant observables, but we will focus here, as previously mentioned,  on the case $ N \ge  k $. For the unitary group the matrix elements $M_{ij}$ are isomorphic to $V \otimes V^*$, where $V^*$ is the complex conjugate representation of $V$. In other words, $U \in U(N)$ acts on $M$ by conjugation,
\begin{equation}
	M \mapsto UMU^\dagger.
\end{equation}
Since $(V \otimes V^*)^{\otimes k} \cong V^{\otimes k} \otimes (V^*)^{\otimes k}$, we have
\begin{equation}
	(V \otimes V^*)^{\otimes k} \cong \qty(\bigoplus_{\Lambda \vdash k} V^{U(N)}_{\Lambda } \otimes V^{\mathbb{C}S_k}_{\Lambda}) \otimes \qty(\bigoplus_{\Lambda' \vdash k} (V^*)^{U(N)}_{\Lambda'} \otimes V^{\mathbb{C}S_k}_{\Lambda'}).
\end{equation}
$U(N)$ invariants appear in a tensor product $V_{\Lambda}^{U(N)} \otimes (V^*)_{\Lambda'}^{U(N)}$ (with multiplicity 1) if and only if $\Lambda = \Lambda'$:
\begin{equation}
	\Dim [V_{\Lambda}^{U(N)} \otimes (V^*)_{\Lambda'}^{U(N)}]^{U(N)} = \delta_{\Lambda \Lambda'}. \label{eq: schur lemma U(N) multiplicity of invariants}
\end{equation}
We are using  $ [W]^{ U(N) } $ to  refer to the $U(N)$ invariant subspace of the representation $W$. We have 
\begin{equation}
	\begin{aligned}
		{[(V \otimes V^*)^{\otimes k}]}^{U(N)} &\cong \bigoplus_{\Lambda, \Lambda' \vdash k} [V_{\Lambda}^{U(N)} \otimes (V^*)_{\Lambda'}^{U(N)}]^{U(N)} \otimes V_{\Lambda}^{\mathbb{C}S_k} \otimes V_{\Lambda'}^{\mathbb{C}S_k} \\
		&\cong \bigoplus_{\Lambda, \Lambda' \vdash k} \delta_{\Lambda \Lambda'} V_{\Lambda}^{\mathbb{C}S_k} \otimes V_{\Lambda'}^{\mathbb{C}S_k} \\
		&\cong  \bigoplus_{\Lambda \vdash k}  V_{\Lambda}^{\mathbb{C}S_k} \otimes V_{\Lambda}^{\mathbb{C}S_k},
	\end{aligned}
\end{equation}
where the second line follows from Schur's Lemma which implies  equation \eqref{eq: schur lemma U(N) multiplicity of invariants}.  Since we are looking for $U(N)$ invariant polynomials of degree $k$ in $M_{ ij}$, the counting is given by the $U(N)$ invariant subspace of $ \Sym^{ k } ( V_N \otimes V_N^* ) $. Equivalently this is the space 
$ {[ ( V_N \otimes V_N^*)^{ \otimes k }  ]^{ U(N) \times S_k }} $. There is a one-dimensional space of $S_k$ invariants in 
$V_{\Lambda}^{\mathbb{C}S_k} \otimes V_{\Lambda}^{\mathbb{C}S_k}$ for each $ \Lambda$. Hence the counting is given by 
\bea \label{eq: U(N) counting}
&& \hbox { Dimension of the space of $U(N)$ invariant  polynomials of degree $k$ in $M_{ij} $ } \cr 
&& =  \sum_{ \Lambda \vdash k } 1 \cr 
&& = \hbox{ Number of integer  partitions of $k$ } \cr 
&& = \hbox{ Number of multi-trace structures with $k$ copies of $M$} .  
\eea
Thus the counting of $U(N)$ invariants is  controlled by the symmetric group algebra, which appeared through Schur-Weyl duality.

Similarly in the case of $S_N$ invariant observables there is a dual algebra at play. The dual algebra for the defining representation of $S_N$ is called the partition algebra, denoted $P_k(N)$ \cite{Martin1994, Martin1996}. The representations of the partition algebra determine the multiplicities of $S_N$ irreducible representations in the decomposition (see section 2.5 in \cite{Halverson2018})
\begin{equation}
	V_N^{\otimes k} \cong \bigoplus_{l=0}^{k}\bigoplus_{\substack{\Lambda_1^\# \vdash l}} V_{\qty[N-l,\Lambda_1^\#]}^{S_N} \otimes V_{\qty[N-l,\Lambda_1^\#]}^{P_k(N)}. \label{eq: Vn otimes k}
\end{equation}
The Young diagram $\Lambda_1 = [N-l,\Lambda_1^\#]$, which is an integer partition of $N$, is constructed by placing the diagram $\Lambda_1^\#$ below a first row of $N-l$ boxes. Requiring $\Lambda_1 $ to be a valid Young diagram imposes some constraints on $ \Lambda_1^{\#} $, which are not manifest in \eqref{eq: Vn otimes k}. This occurs for $ N < 2 k $ as we explain, while it does not occur for $ N \ge 2k$. The latter is called the stable limit. To understand this, we  denote  the first row length of $ \Lambda_1^{ \# } $ 
by $r_1 ( \Lambda_1^{ \# } ) $. 
For $ N \geq 2k $, all values of $l$ and all choices of $\Lambda_1^{ \#  } $ give valid Young diagrams $\Lambda_1 $, since $ N - l \ge r_1 ( \Lambda_1^{ \# } )  $.  Indeed writing $ N = 2k + a $ for $ a \ge 0$, we have 
\bea 
N - l = 2k + a - l \ge k + a . 
\eea
The inequality follows since $ l \le k$ in equation \eqref{eq: Vn otimes k}.  We also have 
\bea 
k + a \ge r_1 (  \Lambda_1^{ \# } ) . 
\eea
This follows because $\Lambda_1^{ \#} $ has no more than $k$ boxes.  For $N < 2k$, the condition 
$ N - l \ge r_1 ( \Lambda_1^{ \# } ) $ imposes a non-trivial $N$-dependent restriction on $ \Lambda_1^{ \# } $. Indeed let $ N  = 2k - a $ for $ a >0  $, then the condition $ N - l \ge r_1 ( \Lambda^{ \# }_1)  $ becomes 
\bea 
k - a \ge   r_1 (  \Lambda_1^{ \# } ) . 
\eea
This is non-trivial condition since $\Lambda_1^{ \# } $ can have up to $k$ boxes.

Note that the symmetric group algebra $\mathbb{C}S_k$ is a subalgebra of $P_k(N)$ (permutations of the tensor factors commute with the action of $S_N$ on $V_N^{\otimes k}$). We can restrict any representation $V_{\Lambda_1}^{P_k(N)}$ to $\mathbb{C}S_k$ to give a decomposition of the form
\begin{equation}
	V_{\Lambda_1}^{P_k(N)} \cong \bigoplus_{\Lambda_2 \vdash k} V_{\Lambda_2}^{S_k} \otimes V_{\Lambda_1 \Lambda_2}^{P_k(N) \rightarrow \mathbb{C}S_k}.
\end{equation}
The dimension of $V_{\Lambda_1 \Lambda_2}^{P_k(N) \rightarrow \mathbb{C}S_k}$ is the branching multiplicity
\begin{align}
\text{Dim} \Big( V_{\Lambda_1 \Lambda_2}^{P_k(N) \rightarrow \mathbb{C}S_k} \Big) = \text{Mult} \Big( V^{P_k(N)}_{\Lambda_1} \rightarrow V^{\mathbb{C}(S_k)}_{\Lambda_2} \Big).
\end{align}
Since $(V_N \otimes V_N)^{\otimes k} \cong V_N^{\otimes k} \otimes V_N^{\otimes k}$ we have
\begin{equation}
	(V_N \otimes V_N)^{\otimes k} \cong  \qty(\bigoplus_{\substack{\Lambda_1\vdash N \\ \Lambda_2 \vdash k }} V_{\Lambda_1}^{S_N} \otimes V_{\Lambda_2}^{S_k} \otimes V_{\Lambda_1 \Lambda_2}^{P_k(N) \rightarrow \mathbb{C}S_k}) \otimes \qty(\bigoplus_{\substack{\Lambda_1' \vdash N \\ \Lambda_2' \vdash k }} V_{\Lambda_1'}^{S_N} \otimes V_{\Lambda_2'}^{S_k} \otimes V_{\Lambda_1' \Lambda_2'}^{P_k(N) \rightarrow \mathbb{C}S_k}).
\end{equation}
There is a single $S_N$ invariant state in every tensor product $V_{\Lambda_1} \otimes V_{\Lambda_1'}$ if and only if $V_{\Lambda_1} \cong V_{\Lambda_1'}$, and similarly for $S_k$. Therefore
\begin{equation}
	[(V_N \otimes V_N)^{\otimes k}]^{S_N \times S_k} \cong 	\bigoplus_{\substack{\Lambda_1 \vdash N \\ \Lambda_2 \vdash k}} V_{\Lambda_1 \Lambda_2}^{P_k(N) \rightarrow \mathbb{C}S_k} \otimes V_{\Lambda_1 \Lambda_2}^{P_k(N) \rightarrow \mathbb{C}S_k}
\end{equation}
and considering the dimension of this subspace of $S_N \times S_k$ invariants, $V_{[N] [k]}$, we find 
\begin{align}\label{sumsqcount} 
	\Dim V_{[N] [k]} = \mathcal{N}(N,k) = \sum_{\Lambda_1 \vdash N} \sum_{\Lambda_1 \vdash k} \text{Mult} 		\Big( V^{P_k(N)}_{\Lambda_1} \rightarrow V^{\mathbb{C}(S_k)}_{\Lambda_2} \Big)^2.
\end{align}
The sum of squares is  indicative of a matrix (Artin-Wedderburn) decomposition \cite{wedderburn1908hypercomplex,artin1927theorie} of a hidden algebra parametrising PIMOs (we found the exposition of the Artin-Wedderburn decomposition in \cite{AR90DissertCh1} to be useful). 
We will turn to an explicit construction of 
PIMOs using partition algebra elements in line with the counting  \eqref{sumsqcount} in section  \ref{sec: PIMOs}.  This sum of squares form in counting invariants, and their connection to the Artin-Wedderburn structure of algebras, have been used in a number of multi-matrix and tensor model applications, e.g. \cite{Kimura2014,PCA2016,Kimura2017,JBSR2017,JBSR2020}.

\subsection{Enhanced $O(N)$ symmetry in parameter space}
\label{subsec: Classical Matrix Model Moments from PIGMM}
The quadratic GOE (Gaussian Orthogonal Ensemble) is determined by the probability density function
\begin{equation}
	\exp({-S(M)}) = \exp(-\Tr( MM^T) ),
\end{equation}
on the space of real symmetric matrices (see definition 2.3.1. in \cite{mehta2004RMT}). The matrix elements $M_{ij}$ for $i \leq j$ in this ensemble of matrices are statistically independent. There are no mixing terms.
Here we consider the underlying space to be the space of real matrices, with no symmetry constraint. There is a 4-parameter family of $O(N)$ invariant quadratic actions
\begin{equation} 
	S(M) = N \epsilon \text{Tr} (M) - \Big(  N \alpha \Tr (MM^T)  + N \beta \Tr(MM) + \gamma ( \Tr M)^2 \Big). \label{eq: general O(N) action}
\end{equation}
In this  model, the matrix elements are not statistically independent, but the linear and quadratic moments are readily solvable, as we now show. Higher moments can be obtained using Wick's theorem. 

This  4-parameter family is a  special case of recently studied \cite{LMT, PIGMM, Ramgoolam2019, PIG2MM} more general Gaussian matrix models, with permutation symmetry.
We now solve for the second moments of matrix variables for the model in 
 \eqref{eq: general O(N) action} and compare with the second moments for the permutation invariant Gaussian 1-matrix model.
This gives a system of linear equations for the parameters in the permutation invariant Gaussian 1-matrix model in terms of the parameters $\alpha,\beta,\gamma$.
See appendix \ref{apx: Param Limit Sage Code} for an algorithm and  computer code to reproduce these results.

We begin by rewriting the action: 
\bea 
&& S  ( M ) = N \epsilon \sum_{ i } M_{ ii}  - N ( \alpha + \beta ) \sum_{ i } M_{ ii}^2 
- N \alpha \sum_{ i \ne j } M_{ij}^2  - N \beta \sum_{ i \ne j } M_{ ij} M_{ ji} 
- \gamma \sum_{ i , j } M_{ ii} M_{ jj}  \, .  \cr 
&& 
\eea
Let 
\begin{align}
	z = (M_{11}, M_{22}, \dots, M_{NN}, M_{12}, M_{21}, M_{13}, M_{31}, \dots,M_{N-1 N}, M_{NN-1}), 
\end{align}
then the action can be expressed as 
\begin{equation}
	S(z) = z \mu - z G z^T . 
\end{equation}
The vector $\mu$ is 
\begin{align}
	\mu = \mqty(N \epsilon \\ \vdots \\ N \epsilon \\ 0 \\ \vdots \\ 0)
\end{align}
with the first $N$ terms equal to $\epsilon$ and the rest $0$ and
\begin{equation}
	\begin{aligned}
		G  = 
		\mqty(\dmat{G_1, G_2, \ddots, G_2}), \quad &G_1 = N \mqty(\dmat{\alpha + \beta, \ddots, \alpha + \beta}) + \mqty(\gamma & \dots & \gamma \\ \vdots & \vdots & \vdots \\ \gamma & \dots & \gamma), \\
		&G_2 = N \mqty(\alpha & \beta \\ \beta & \alpha).
	\end{aligned}
\end{equation}
The inverse of $G_2$ is 
\begin{align}
	\big( G_2 \big)^{-1} = \frac{1}{N \big( \alpha^2 - \beta^2 \big)}
	\mqty(\alpha & -\beta \\ -\beta & \alpha)
\end{align}
while the inverse of $G_1$ is given by
\begin{align}
	\big( G_1 \big)^{-1}_{ij} = 
	\begin{cases}
		\frac{1}{N^2} \Big( \frac{N-1}{\alpha + \beta} + \frac{1}{\alpha + \beta + \gamma} \Big) ,& \text{if } i = j \\
		- \frac{1}{N^2} \Big( \frac{\gamma}{(\alpha + \beta)(\alpha + \beta + \gamma)} \Big) ,& \text{if } i \neq j 
	\end{cases}
\end{align}
From the form of these inverse matrices we can write down the connected two-point function
\begin{align} \nonumber
	\langle \tensor{M}{_i_j} \tensor{M}{_k_l} \rangle_{{\con}} &= \tensor{\delta}{_i_j} \tensor{\delta}{_k_l} \tensor{\delta}{_i_l} \frac{1}{N^2} \Big( \frac{N-1}{\alpha + \beta} + \frac{1}{\alpha + \beta + \gamma}  \Big) - \big( \tensor{\delta}{_i_j} \tensor{\delta}{_k_l} - \tensor{\delta}{_i_j} \tensor{\delta}{_k_l} \tensor{\delta}{_i_l} \big) \frac{1}{N^2} \frac{\gamma}{(\alpha + \beta)(\alpha + \beta + \gamma)} \\
	&+ \big( \tensor{\delta}{_i_k} \tensor{\delta}{_j_l} - \tensor{\delta}{_i_k} \tensor{\delta}{_j_l} \tensor{\delta}{_i_j} \big) \frac{1}{N} \frac{\alpha}{\alpha^2 - \beta^2} - \big( \tensor{\delta}{_i_l} \tensor{\delta}{_k_j} - \tensor{\delta}{_i_l} \tensor{\delta}{_k_j} \tensor{\delta}{_i_j} \big) \frac{1}{N} \frac{\beta}{\alpha^2 - \beta^2}.
\end{align}
Defining
\begin{align} \nonumber
	a &= \frac{1}{N^2} \Big( \frac{N-1}{\alpha + \beta} + \frac{1}{\alpha + \beta + \gamma}  \Big), \qquad &&b = \frac{1}{N^2} \frac{\gamma}{(\alpha + \beta)(\alpha + \beta + \gamma)} \\
	c &= \frac{1}{N} \frac{\alpha}{\alpha^2 - \beta^2}, \qquad &&d = \frac{1}{N} \frac{\beta}{\alpha^2 - \beta^2}
\end{align}
and collecting like terms we are left with the following expression for the two-point function
\begin{align}
	\langle  \tensor{M}{_i_j} \tensor{M}{_k_l} \rangle_{\con} = \tensor{\delta}{_i_j} \tensor{\delta}{_k_l} \tensor{\delta}{_i_l} \big( a+b-c+d \big) -  \tensor{\delta}{_i_j} \tensor{\delta}{_k_l} b  +  \tensor{\delta}{_i_k} \tensor{\delta}{_j_l} c -  \tensor{\delta}{_i_l} \tensor{\delta}{_k_j} d.
\end{align}
The parameters $a,b,c,d$ satisfy $a+b+d=c$ and therefore the fully simplified two-point function is given by
\begin{equation}
	\langle  \tensor{M}{_i_j} \tensor{M}{_k_l} \rangle_{\con} = -\tensor{\delta}{_i_j} \tensor{\delta}{_k_l} b  +  \tensor{\delta}{_i_k} \tensor{\delta}{_j_l}(a+b+d) -  \tensor{\delta}{_i_l} \tensor{\delta}{_k_j} d.
\end{equation}
Comparing this to the two-point function of the permutation invariant matrix model (equation (3.6) in \cite{PIGMM}) we find that it is reproduced in following parameter limit
\begin{align}\label{eqn:5in13} \nonumber
	&\big( g_{[N]}^{-1} \big)_{11} = a \\ \nonumber 
	&\big( g_{[N]}^{-1} \big)_{22} = a-(N-2)b\\ \nonumber
	&\big( g_{[N]}^{-1} \big)_{12} = -\sqrt{N-1}b\\ \nonumber
	&\big( g_{[N-1,1]}^{-1} \big)_{11} = a+b+d \\ \nonumber
	&\big( g_{[N-1,1]}^{-1} \big)_{22} = a+b+d\\ \nonumber
	&\big( g_{[N-1,1]}^{-1} \big)_{33} =  a+b\\ \nonumber
	&\big( g_{[N-1,1]}^{-1} \big)_{12} = -d\\ \nonumber
	&\big( g_{[N-1,1]}^{-1} \big)_{13} = 0\\ \nonumber
	&\big( g_{[N-1,1]}^{-1} \big)_{23} =  0\\ \nonumber
	&\big( g_{[N-2,2]}^{-1} \big) = a+b \\
	&\big( g_{[N-2,1,1]}^{-1} \big) = a+b+2d
\end{align}
where we have again written $g$ instead of $\Lambda$ for our quadratic couplings, labelling them using integer partitions under the identification given in \eqref{eq: PIGMM irrep labels}.

There is a special point in this limit that recovers the two-point function for the simple $O(N)$ model with action
\begin{align}\label{eqn:simplestON} 
	S(M) =  \Tr( M M^T ).
\end{align}
Setting $\epsilon = \beta = \gamma = 0$ in equation \eqref{eq: general O(N) action}, we find that the relevant limit of the permutation invariant Gaussian model is found by taking $a = 1$ and $b = d = 0$ in \eqref{eqn:5in13} which gives us
\begin{align} \label{eqn:1in13}
	\big( g_{[N]}^{-1} \big)_{11} = \big( g_{[N]}^{-1} \big)_{22} =
	\big( g_{[N-1,1]}^{-1} \big)_{11} = \big( g_{[N-1,1]}^{-1} \big)_{22} = \big( g_{[N-1,1]}^{-1} \big)_{33} =
	\big( g_{[N-2,2]}^{-1} \big) = \big( g_{[N-2,1,1]}^{-1} \big) = 1
\end{align}
as the only non-zero parameters.

A quick check on the above computation is the following. Using Clebsch-Gordan coefficients we have
\begin{equation}
	\begin{aligned}		
	\Tr(MM^T) = \sum_{ij} M_{ij}M_{ij} &= \sum_{ij} \sum_{a,b, \Lambda_1 \Lambda_1', \alpha, \beta}C_{a,ij}^{\Lambda_1, \alpha}C_{b,ij}^{\Lambda_1', \beta}S_{a}^{\Lambda_1, \alpha}S_{b}^{\Lambda_1', \beta} \\
	&=  \sum_{a,b, \Lambda_1 \Lambda_1', \alpha, \beta} \delta_{ab}\delta^{\Lambda_1 \Lambda_1'}\delta^{\alpha \beta} S_{a}^{\Lambda_1, \alpha}S_{b}^{\Lambda_1', \beta} 	= \sum_{a,\Lambda_1, \alpha}S_{a}^{\Lambda_1, \alpha} S_{a}^{\Lambda_1, \alpha},
	\end{aligned} 
\end{equation}
where the second line uses orthogonality of the Clebsch-Gordan coefficients.
Comparing with equation \eqref{eq: SN partition fn} recovers the parameter limit \eqref{eqn:1in13}.\footnote{Immediate comparison gives a parameter limit for the coupling matrices, as opposed to their inverses as in equation \eqref{eqn:1in13}. In this case, they are identical.}

\section{Permutation Invariant Matrix Observables (PIMOs) }
\label{sec: PIMOs}
We will now describe the partition algebra and how the PIMOs are constructed from the $S_k$ invariant subalgebra of $P_k(N)$.
Properties of the partition algebra \cite{Jones1994, Martin1994, Martin1996,Halverson2004} will allow us to prove large $N$ factorisation of PIMOs in the $O(N)$ symmetric matrix model.

The partition algebra $P_k(N)$ is a diagram algebra. It has a finite basis, labelled by diagrams, where multiplication is a type of composition of diagrams.
A diagram in $P_k(N)$ has $2k$ labelled vertices arranged into two rows, with $k$ vertices in each row. Any set of edges between the vertices are allowed.
We use the convention in which the bottom vertices are labelled (from left to right) by $1, \dots,k$ and the top vertices by $1', \dots, k'$. For example, $P_2(N)$ has a basis of $15$ diagrams. Among these are,
\begin{align}
	&\PAdiagramLabeled{2}{-1/-2}, \quad\PAdiagramLabeled{2}{-1/-2,1/2} \\
	&\PAdiagramLabeled{2}{1/-2,2/-1}, \quad \PAdiagramLabeled{2}{1/2,2/-2}.
\end{align}
In general, the dimension of $P_k(N)$ is the number of set partitions of $2k$ (also known as Bell numbers).

The underlying set for this basis of the partition algebra is the set of set partitions of the $2k$ labelled vertices.
There is a redundancy in the diagram picture.
The redundancy arises from the fact that we are free to choose any set of edges, as long as every vertex in a subset of the set partition can be reached from any other vertex in the same subset, by a path along the edges.
For example, the following pair of diagrams correspond to the same element.
\begin{equation}
	\PAdiagram{3}{2/3,3/-3} = \PAdiagram{3}{2/-3,3/-3}.
\end{equation}
The product in $P_k(N)$ is independent of the choice of representative diagram.

Let $d_1$ and $d_2$ be two diagrams in $P_k(N)$.
The composition $d_3 = d_1 d_2$ is constructed by placing $d_1$ above $d_2$ and identifying the bottom vertices of $d_1$ with the top vertices of $d_2$. The diagram is simplified by following the edges connecting the bottom vertices of $d_2$ to the top vertices of $d_1$. Any connected components within  the middle rows are removed and we multiply by $N^c$, where $c$ is the number of these components removed.
For example,
\begin{equation}
	\begin{aligned}
		&\PAdiagram{3}{-1/-2,-3/2} \\
		&\PAdiagram{3}{1/2,-2/-3,-3/3}
	\end{aligned} = N \PAdiagram{3}{-2/-3,-3/2}
	\qq{and} 
	\begin{aligned}
		&\PAdiagram{3}{-1/1,-2/3,-3/2} \\
		&\PAdiagram{3}{-1/2,-2/-3,-3/3}
	\end{aligned} = \PAdiagram{3}{-1/3,-2/-3,-3/2},
\end{equation}
where the factor of $N$ in the first equation comes from removing the middle component at vertex $1$ and $2$.
For linear combinations of diagrams, multiplication is defined by linear extension.

The subset of diagrams with $k$ edges, each connecting a vertex at the top to a vertex at the bottom and where every vertex has exactly one incident edge, span a subalgebra. This subalgebra is isomorphic to the symmetric group algebra $\mathbb{C}S_k$. For example, there is a one-to-one correspondence between permutations in $S_3$ and the following set of diagrams,
\begin{equation}
	\PAdiagram{3}{1/-1,2/-2,3/-3}, \PAdiagram{3}{1/-2,2/-1,3/-3}, \PAdiagram{3}{1/-3,2/-2,3/-1}, \PAdiagram{3}{1/-1,2/-3,3/-2}, \PAdiagram{3}{1/-3,2/-1,3/-2}, \PAdiagram{3}{1/-2,2/-3,3/-1}.
\end{equation}

\subsection{Construction of PIMOs}
We will construct degree $k$ PIMOs from elements $d \in P_k(N)$. As a warm-up, we recap the construction of $U(N)$ invariants using elements in $\mathbb{C}S_k$. See \cite{Ramgoolam2016} for a review of the background literature.

For this construction it will be useful to rewrite $M_{ij}$ as $M^i_j$ and think of these as the matrix elements of a linear operator acting on $V$, the defining representation of $U(N)$.
Define $M$ to be the linear operator $M: V \rightarrow V$ with matrix elements
\begin{equation}
	Me_{i} = \sum_{j} M^{j}_i e_{j} ,
\end{equation}
in a basis $e_i$ for $V$.
In diagram notation the linear operator $M$ is represented by a box labelled $M$, with one incoming and one outgoing  edge,
\begin{equation}
	M^{j}_i = \begin{tikzpicture}[baseline=(base)]
		\coordinate (base) at (0,0);
		\node[label=below: {$i$},circle,fill,inner sep=1pt] (b1) at (0 ,-.5) {};
		\node[label=above: {$j$},circle,fill,inner sep=1pt] (t1) at (0 ,.5) {};		
		\draw[fill=white] (-.30,.25) rectangle node{$M$} (.30,-.25);
		\draw (0,-.25) -- (b1);
		\draw (0,.25) -- (t1); 	
	\end{tikzpicture}
\end{equation}
The operator $M^{\otimes k}$ acts on $V^{\otimes k}$ as
\begin{equation}
	M^{\otimes k} e_{i_1} \otimes \dots \otimes e_{i_k} = Me_{i_1} \otimes \dots \otimes Me_{i_k}.
\end{equation}
Diagrammatically, tensor products of operators are represented by horizontally composing the diagrams,
\begin{equation}
	(M^{\otimes k})^{j_1 \dots j_k}_{i_1 \dots i_k} = M^{j_1}_{i_1} \dots M^{j_k}_{i_k} = \begin{tikzpicture}[baseline=(base)]
		\coordinate (base) at (0,0); 
		\node[label=below: {$i_1$},circle,fill,inner sep=1pt] (b1) at (-.7 ,-.5) {};
		\node[label=below: {$i_k$},circle,fill,inner sep=1pt] (bk) at (.7 ,-.5) {};
		\node[label=above: {$j_{1}$},circle,fill,inner sep=1pt] (t1) at (-.7 ,.5) {};
		\node[label=above: {$j_k$},circle,fill,inner sep=1pt] (tk) at (.7 ,.5) {};	
		\draw[loosely dotted] (-.4,0) -- (.4,0);	
		\draw (-.7,-.25) -- (b1);	\draw (.7,-.25) -- (bk);
		\draw (-.7,.25) -- (t1); 	\draw (.7,.25) -- (tk);	
		\draw[fill=white] (-1.1,.25) rectangle node{$M$} (-.5,-.25);
		\draw[fill=white] (.5,.25) rectangle node{$M$} (1.1,-.25);
	\end{tikzpicture}.
\end{equation}

When viewed as a matrix polynomial, the trace
\begin{equation}
	\mathcal{O}_{\tau} = \Tr_{V^{\otimes k}} (M^{\otimes k} \tau) = \sum_{\substack{i_1 \dots i_k \\ i_{1'} \dots i_{k'}}}(\tau)^{i_{1'} \dots i_{k'}}_{i_1 \dots i_k}M^{i_1}_{i_{1'}} \dots M^{i_k}_{i_{k'}}=\begin{tikzpicture}[baseline=(base)]
		\coordinate (base) at (0,.5); 
		\coordinate (b1) at (-.8 ,-.5) {};
		\coordinate (bk) at (.8 ,-.5) {};
		\coordinate (t1) at (-.8 ,1.5) {};
		\coordinate (tk) at (.8,1.5) {};	
		\draw[loosely dotted] (-.4,0) -- (.4,0);	
		\draw (-.8,-.25) -- (b1);	\draw (.8,-.25) -- (bk);
		\draw (-.8,.25) -- (t1); 	\draw (.8,.25) -- (tk);
		\draw ($(b1)-(0.2,0)$ ) -- ($(bk)+(0.2,0)$);
		\draw ($(t1)-(0.2,0)$) -- ($(tk)+(0.2,0)$);
		\draw[fill=white] (-1.1,1) rectangle node{$\tau$} (1.1,0.5);	
		\draw[fill=white] (-1.1,.25) rectangle node{$M$} (-.5,-.25);
		\draw[fill=white] (.5,.25) rectangle node{$M$} (1.1,-.25);
	\end{tikzpicture}, \label{eq: UN operator as diagram}
\end{equation}
is a unitary invariant of degree $k$. The matrix elements of the permutation $\tau$ as a linear operator on $V^{\otimes k}$ are
\begin{equation}
	(\tau)^{i_{1'} \dots i_{k'}}_{i_1 \dots i_k} = \delta^{i_{1'}}_{i_{\tau(1)}} \dots \delta^{i_{k'}}_{i_{\tau(k)}}.
\end{equation}
The diagram representing $\tau$ is obtained by associating an edge with every Kronecker delta.
For example, for $\tau = (12)$ we have the diagram
\begin{equation}
	\begin{tikzpicture}[baseline=(base)]
		\coordinate (base) at (0,0 ); 
		\node[label=below: {$i_1$},circle,fill,inner sep=1pt] (b1) at (-.5 ,-.5) {};
		\node[label=below: {$i_2$},circle,fill,inner sep=1pt] (bk) at (.5 ,-.5) {};
		\node[label=above: {$i_{1'}$},circle,fill,inner sep=1pt] (t1) at (-.5 ,.5) {};
		\node[label=above: {$i_{2'}$},circle,fill,inner sep=1pt] (tk) at (.5 ,.5) {};
		\draw (-.5,-.25) -- (.5,.25);
		\draw (.5,-.25) -- (-.5,.25);
		\draw (-.5,-.25) -- (b1);	\draw (.5,-.25) -- (bk);
		\draw (-.5,.25) -- (t1); 	\draw (.5,.25) -- (tk);	
	\end{tikzpicture} = \delta^{i_{2'}}_{i_1}\delta^{i_{1'}}_{i_2}.
\end{equation}
The horizontal lines in equation \eqref{eq: UN operator as diagram} are used to indicate that the incoming and outgoing edges are identified, as expected from a trace.

Invariance under the action of $U(N)$ follows because $\tau \in S_k$ commutes with any $U(N)$ acting on $V^{\otimes k}$.
The correspondence between gauge invariant operators and permutations has a redundancy given by,
\begin{equation}
	\mathcal{O}_{\gamma \tau \gamma^{{-1}}} = \mathcal{O}_{\tau}, \qq{for all $\gamma \in S_k$.}
\end{equation}
This follows because $\gamma^{-1}M^{\otimes k}\gamma = M^{\otimes k}$.
Therefore, a basis of multi-trace observables is in one-to-one correspondence with conjugacy classes of $S_k$, as expected from the counting in equation \eqref{eq: U(N) counting}.

The construction of degree $k$ PIMOs from elements of $P_k(N)$ is identical.
For any $d \in P_k(N)$, the matrix polynomial
\begin{equation}
	\mathcal{O}_{d} = \Tr_{V_N^{\otimes k}} (M^{\otimes k} d) = \sum_{\substack{i_1 \dots i_k \\ i_{1'} \dots i_{k'}}}(d)^{i_{1'} \dots i_{k'}}_{i_1 \dots i_k}M^{i_1}_{i_{1'}} \dots M^{i_k}_{i_{k'}}=\begin{tikzpicture}[baseline=(base)]
		\coordinate (base) at (0,.5); 
		\coordinate (b1) at (-.8 ,-.5) {};
		\coordinate (bk) at (.8 ,-.5) {};
		\coordinate (t1) at (-.8 ,1.5) {};
		\coordinate (tk) at (.8,1.5) {};	
		\draw[loosely dotted] (-.4,0) -- (.4,0);	
		\draw (-.8,-.25) -- (b1);	\draw (.8,-.25) -- (bk);
		\draw (-.8,.25) -- (t1); 	\draw (.8,.25) -- (tk);
		\draw ($(b1)-(0.2,0)$ ) -- ($(bk)+(0.2,0)$);
		\draw ($(t1)-(0.2,0)$) -- ($(tk)+(0.2,0)$);
		\draw[fill=white] (-1.1,1) rectangle node{$d$} (1.1,0.5);	
		\draw[fill=white] (-1.1,.25) rectangle node{$M$} (-.5,-.25);
		\draw[fill=white] (.5,.25) rectangle node{$M$} (1.1,-.25);
	\end{tikzpicture}, \label{eq: PIMO from PkN}
\end{equation}
is a PIMO, because $d$ commutes with the action of $S_N$ acting on $V_N^{\otimes k}$. The matrix elements $(d)^{i_{1'} \dots i_{k'}}_{i_1 \dots i_k}$ also correspond to 
 the diagram representation by associating every Kronecker delta to an  edge connecting a pair of vertices. For example,
\begin{equation}
	\PAdiagramLabeled{2}{-1/-2,1/2,-1/1,-2/2} = \delta_{i_1 i_2} \delta_{i_2}^{i_{2'}} \delta^{i_{2'} i_{1'}} \quad \text{and} \quad \PAdiagramLabeled{2}{-1/1,-1/-2} = \delta_{i_1 i_2} \delta_{i_1}^{i_{1'}}. \label{eq: examples of diagram tensors}
\end{equation}
As before, for any $\gamma \in S_k$ we have
\begin{equation}
	\mathcal{O}_{\gamma d\gamma^{-1}} =\mathcal{O}_{d}.
\end{equation}
Degree $k$ PIMOs are in one-to-one correspondence with the $S_k$ invariant subalgebra of $P_k(N)$. A basis is given by the set of distinct equivalence classes
\begin{equation}\label{Equivs} 
	[d] = \{\gamma d \gamma^{-1} \, \vert \, \forall \gamma \in S_k\}.
\end{equation}

\subsection{Inner product on PIMOs}
The simplest $O(N)$ invariant matrix model has the quadratic expectation value 
\begin{equation} \label{eq: O(N) two-point function}
	\langle M^{i}_j M^{k}_{l} \rangle = \delta^{ik}\delta_{jl}.
\end{equation}
Let $d_1, d_2 \in P_k(N)$, and define the two-point function of PIMOs $ \cO_{ d_1} , \cO_{ d_2}$ by using Wick's theorem and the equation \eqref{eq: O(N) two-point function}, keeping only Wick contractions between the two observables i.e. we are treating them as ``normal-ordered''. 
There is an expression for this two-point function 
\begin{equation}
	\expval{\normord{\mathcal{O}_{d_1}}\normord{\mathcal{O}_{d_2}}} = \sum_{\gamma \in S_k} \Tr_{V_N^{\otimes k}}(d_1 \gamma d_2^T \gamma^{-1}), \label{eq: Two Point Function of Partition Algebra Observables}
\end{equation} 
where $d^T$ is the transpose of the diagram $d$, obtained from $d$ by flipping the top and bottom vertices. The permutations $\gamma$ parameterize the Wick contractions. The proof of \eqref{eq: Two Point Function of Partition Algebra Observables} goes as follows.
Note that the quadratic expectation value \eqref{eq: O(N) two-point function} diagrammatically corresponds to the replacement
\begin{equation}
	\expval{\begin{tikzpicture}[baseline=(base)]
			\coordinate (base) at (0,0);
			\node[label=below: {$j$},circle,fill,inner sep=1pt] (b1) at (0 ,-.5) {};
			\node[label=above: {$i$},circle,fill,inner sep=1pt] (t1) at (0 ,.5) {};		
			\draw[fill=white] (-.30,.25) rectangle node{$M$} (.30,-.25);
			\draw (0,-.25) -- (b1);
			\draw (0,.25) -- (t1); 	
		\end{tikzpicture} \begin{tikzpicture}[baseline=(base)]
			\coordinate (base) at (0,0);
			\node[label=below: {$l$},circle,fill,inner sep=1pt] (b1) at (0 ,-.5) {};
			\node[label=above: {$k$},circle,fill,inner sep=1pt] (t1) at (0 ,.5) {};		
			\draw[fill=white] (-.30,.25) rectangle node{$M$} (.30,-.25);
			\draw (0,-.25) -- (b1);
			\draw (0,.25) -- (t1); 	
	\end{tikzpicture}} = \begin{tikzpicture}[baseline=(base)]
		\coordinate (base) at (0,0 ); 
		\node[label=below: {$j$},circle,fill,inner sep=1pt] (b1) at (-.5 ,-.5) {};
		\node[label=below: {$l$},circle,fill,inner sep=1pt] (bk) at (.5 ,-.5) {};
		\node[label=above: {$i$},circle,fill,inner sep=1pt] (t1) at (-.5 ,.5) {};
		\node[label=above: {$k$},circle,fill,inner sep=1pt] (tk) at (.5 ,.5) {};
		\draw (.5,-.25) -- (-.5,-.25);
		\draw (.5,.25) -- (-.5,.25);
		\draw (-.5,-.25) -- (b1);	\draw (.5,-.25) -- (bk);
		\draw (-.5,.25) -- (t1); 	\draw (.5,.25) -- (tk);	
	\end{tikzpicture},
\end{equation}
where the Kronecker deltas have been replaced by edges. The two-point function in equation \eqref{eq: Two Point Function of Partition Algebra Observables} can be represented by the diagram in the first line below
\begin{equation}
	\begin{aligned}
		\expval{\normord{\mathcal{O}_{d_1}}\normord{\mathcal{O}_{d_2}}}_{\con}
		& = \expval{\begin{tikzpicture}[baseline=(base)]
				\coordinate (base) at (0,.25); 
				\coordinate (b1) at (-.8 ,-.5) {};
				\coordinate (bk) at (.8 ,-.5) {};
				\coordinate (t1) at (-.8 ,1.5) {};
				\coordinate (tk) at (.8,1.5) {};	
				\draw[loosely dotted] (-.4,0) -- (.4,0);	
				\draw (-.8,-.25) -- (b1);	\draw (.8,-.25) -- (bk);
				\draw (-.8,.25) -- (t1); 	\draw (.8,.25) -- (tk);
				\draw ($(b1)-(0.2,0)$ ) -- ($(bk)+(0.2,0)$);
				\draw ($(t1)-(0.2,0)$) -- ($(tk)+(0.2,0)$);
				\draw[fill=white] (-1.1,1) rectangle node{$d_1$} (1.1,0.5);	
				\draw[fill=white] (-1.1,.25) rectangle node{$M$} (-.5,-.25);
				\draw[fill=white] (.5,.25) rectangle node{$M$} (1.1,-.25);
			\end{tikzpicture} \begin{tikzpicture}[baseline=(base)]
				\coordinate (base) at (0,0.25); 
				\coordinate (b1) at (-.8 ,-.5) {};
				\coordinate (bk) at (.8 ,-.5) {};
				\coordinate (t1) at (-.8 ,1.5) {};
				\coordinate (tk) at (.8,1.5) {};	
				\draw[loosely dotted] (-.4,0) -- (.4,0);	
				\draw (-.8,-.25) -- (b1);	\draw (.8,-.25) -- (bk);
				\draw (-.8,.25) -- (t1); 	\draw (.8,.25) -- (tk);
				\draw ($(b1)-(0.2,0)$ ) -- ($(bk)+(0.2,0)$);
				\draw ($(t1)-(0.2,0)$) -- ($(tk)+(0.2,0)$);
				\draw[fill=white] (-1.1,1) rectangle node{$d_2$} (1.1,0.5);	
				\draw[fill=white] (-1.1,.25) rectangle node{$M$} (-.5,-.25);
				\draw[fill=white] (.5,.25) rectangle node{$M$} (1.1,-.25);
		\end{tikzpicture}} \\[1em]
		&= \sum_{\gamma \in S_k} 
		\begin{tikzpicture}[baseline = (m1)]
			\coordinate (base) at (0,-.25); 
			\coordinate (size) at (2,0);
			\coordinate (sep) at (-1,0);
			\coordinate (mar) at (0.2,0);
			\coordinate (i1) at (0,0) {}; \coordinate (i2) at ($(i1)+(size)$) {}; \coordinate (i3) at ($(i2)+(size)+(sep)$) {}; \coordinate (i4) at ($(i3)+(size)$) {};
			\coordinate (j1) at (0,.25) {}; \coordinate (j2) at ($(j1)+(size)$) {}; \coordinate (j3) at ($(j2)+(size)+(sep)$) {}; \coordinate (j4) at ($(j3)+(size)$) {};
			\coordinate (k1) at (0,.75) {}; \coordinate (k2) at ($(k1)+(size)$) {}; \coordinate (k3) at ($(k2)+(size)+(sep)$) {}; \coordinate (k4) at ($(k3)+(size)$) {};
			\coordinate (l1) at (0,1) {}; \coordinate (l2) at ($(l1)+(size)$) {}; \coordinate (l3) at ($(l2)+(size)+(sep)$) {}; \coordinate (l4) at ($(l3)+(size)$) {};
			\coordinate (m1) at (0,1.5) {}; \coordinate (m2) at ($(m1)+(size)$) {}; \coordinate (m3) at ($(m2)+(size)+(sep)$) {}; \coordinate (m4) at ($(m3)+(size)$) {};
			\coordinate (n1) at (0,1.75) {}; \coordinate (n2) at ($(n1)+(size)$) {}; \coordinate (n3) at ($(n2)+(size)+(sep)$) {}; \coordinate (n4) at ($(n3)+(size)$) {};
			\coordinate (o1) at (0,2.25) {}; \coordinate (o2) at ($(o1)+(size)$) {}; \coordinate (o3) at ($(o2)+(size)+(sep)$) {}; \coordinate (o4) at ($(o3)+(size)$) {};
			\coordinate (p1) at (0,2.5) {}; \coordinate (p2) at ($(p1)+(size)$) {}; \coordinate (p3) at ($(p2)+(size)+(sep)$) {}; \coordinate (p4) at ($(p3)+(size)$) {};
			\coordinate (q1) at (0,3) {}; \coordinate (q2) at ($(q1)+(size)$) {}; \coordinate (q3) at ($(q2)+(size)+(sep)$) {}; \coordinate (q4) at ($(q3)+(size)$) {};
			\coordinate (r1) at (0,3.25) {}; \coordinate (r2) at ($(r1)+(size)$) {}; \coordinate (r3) at ($(r2)+(size)+(sep)$) {}; \coordinate (r4) at ($(r3)+(size)$) {};
			\draw (i1) -- (j1) -- (k1) -- (l1) to[out=90, in=90, looseness=0.1] (l3) -- (k3) -- (j3) -- (i3);
			\draw (i2) -- (j2) -- (k2) -- (l2) to[out=90, in=90, looseness=0.1] (l4) -- (k4) -- (j4) -- (i4);
			\draw (r3) -- (m3) to[out=-90, in=-90, looseness=0.1] (m1) -- (r1);
			\draw (r4) -- (m4) to[out=-90, in=-90, looseness=0.1] (m2) -- (r2);
			\draw[fill=white] ($(k1)-(mar)$) rectangle node{$\gamma^{{-1}}$} ($(j2)+(mar)$);	\draw[fill=white] ($(o1)-(mar)$) rectangle node{$\gamma$} ($(n2)+(mar)$);	
			\draw[fill=white] ($(q1)-(mar)$) rectangle node{$d_1$} ($(p2)+(mar)$); \draw[fill=white] ($(q3)-(mar)$) rectangle node{$d_2$} ($(p4)+(mar)$);	
			\draw ($(i1)-(mar)$) -- ($(i2)+(mar)$);	 \draw ($(r1)-(mar)$) -- ($(r2)+(mar)$);
			\draw ($(i3)-(mar)$) -- ($(i4)+(mar)$);	 \draw ($(r3)-(mar)$) -- ($(r4)+(mar)$);
		\end{tikzpicture} \\[1em]
		&= \sum_{\gamma \in S_k}
		\begin{tikzpicture}[baseline = (m1)]
			\coordinate (base) at (0,-.25); 
			\coordinate (size) at (2,0);
			\coordinate (sep) at (0,0);
			\coordinate (mar) at (0.2,0);
			\coordinate (i1) at (0,0) {}; \coordinate (i2) at ($(i1)+(size)$) {}; \coordinate (i3) at ($(i2)+(size)+(sep)$) {}; \coordinate (i4) at ($(i3)+(size)$) {};
			\coordinate (j1) at (0,.25) {}; \coordinate (j2) at ($(j1)+(size)$) {}; \coordinate (j3) at ($(j2)+(size)+(sep)$) {}; \coordinate (j4) at ($(j3)+(size)$) {};
			\coordinate (k1) at (0,.75) {}; \coordinate (k2) at ($(k1)+(size)$) {}; \coordinate (k3) at ($(k2)+(size)+(sep)$) {}; \coordinate (k4) at ($(k3)+(size)$) {};
			\coordinate (l1) at (0,1) {}; \coordinate (l2) at ($(l1)+(size)$) {}; \coordinate (l3) at ($(l2)+(size)+(sep)$) {}; \coordinate (l4) at ($(l3)+(size)$) {};
			\coordinate (m1) at (0,1.5) {}; \coordinate (m2) at ($(m1)+(size)$) {}; \coordinate (m3) at ($(m2)+(size)+(sep)$) {}; \coordinate (m4) at ($(m3)+(size)$) {};
			\coordinate (n1) at (0,1.75) {}; \coordinate (n2) at ($(n1)+(size)$) {}; \coordinate (n3) at ($(n2)+(size)+(sep)$) {}; \coordinate (n4) at ($(n3)+(size)$) {};
			\coordinate (o1) at (0,2.25) {}; \coordinate (o2) at ($(o1)+(size)$) {}; \coordinate (o3) at ($(o2)+(size)+(sep)$) {}; \coordinate (o4) at ($(o3)+(size)$) {};
			\coordinate (p1) at (0,2.5) {}; \coordinate (p2) at ($(p1)+(size)$) {}; \coordinate (p3) at ($(p2)+(size)+(sep)$) {}; \coordinate (p4) at ($(p3)+(size)$) {};
			\coordinate (q1) at (0,3) {}; \coordinate (q2) at ($(q1)+(size)$) {}; \coordinate (q3) at ($(q2)+(size)+(sep)$) {}; \coordinate (q4) at ($(q3)+(size)$) {};
			\coordinate (r1) at (0,3.25) {}; \coordinate (r2) at ($(r1)+(size)$) {}; \coordinate (r3) at ($(r2)+(size)+(sep)$) {}; \coordinate (r4) at ($(r3)+(size)$) {};
			\draw (i1) -- (r1); \draw (i2) -- (r2); 
			\draw[fill=white] ($(k1)-(mar)$) rectangle node{$\gamma^{{-1}}$} ($(j2)+(mar)$);	\draw[fill=white] ($(o1)-(mar)$) rectangle node{$\gamma$} ($(n2)+(mar)$);	
			\draw[fill=white] ($(q1)-(mar)$) rectangle node{$d_1$} ($(p2)+(mar)$); \draw[fill=white] ($(m1)-(mar)+(0,0.1)$) rectangle node{$d_2^T$} ($(l2)+(mar)-(0,0.1)$);	
			\draw ($(i1)-(mar)$) -- ($(i2)+(mar)$);	 \draw ($(r1)-(mar)$) -- ($(r2)+(mar)$);
		\end{tikzpicture}
	\end{aligned}
\end{equation}
The second line is the sum over Wick contractions parameterized by $\gamma \in S_k$.
The last equality comes from straightening the diagram. By following the lines and recording the operators encountered on the way, we recognize the last diagram as the representation of $\Tr_{V_N^{\otimes k}}(d_1 \gamma d_2^T \gamma^{-1})$.

The symmetry of the two-point function is proved by observing that 
\bea\label{Symmetry}   
\sum_{ \gamma  \in S_k } \Tr_{V_N^{\otimes k}}  ( d_1 \gamma d_2^T \gamma^{-1} ) = \sum_{ \gamma \in S_k  } \Tr_{V_N^{\otimes k}} ( \gamma d_2 \gamma^{-1} d_1^T) = \sum_{ \gamma \in S_k  }  \Tr_{V_N^{\otimes k}} ( d_2 \gamma d_1^T \gamma^{-1} ) . 
\eea
We have used the invariance of the trace under transposition, cyclicity of the trace and a relabelling of $ \gamma \rightarrow \gamma^{-1} $. The non-degeneracy of the two-point function  at large $N$ follows from the factorization property in the next section. The non-degeneracy
at all orders in $1/\sqrt{N}$ is proved in the companion paper by exhibiting an orthogonal basis constructed using representation theory data \cite{MQMPartAlg}. This shows that the two-point function defines an inner product. 

\section{Large N factorisation}
\label{sec: factorization}
In this section, we will show that the normalized PIMOs
\begin{equation}
	\mathcal{\hat{O}}_{d} = \frac{\mathcal{O}_{d}}{\sqrt{\expval{\normord{\mathcal{O}_{d}}\normord{\mathcal{O}_{d}}}_{\con}}},
\end{equation}
factorize for large $N$
\begin{equation}
	\expval{\normord{\mathcal{\hat{O}}_{d_1}}\normord{\mathcal{\hat{O}}_{d_2}}}_{\con} = \begin{cases}
		1 + O(1/\sqrt{N}) \qq{if $[d_1] = [d_2]$,} \\
		0 + O(1/\sqrt{N}) \qq{if $[d_1] \neq [d_2]$.}
	\end{cases}
\end{equation}
To prove large $N$ factorization we will study the powers of $N$ appearing in
\begin{equation}
	\Tr_{V_N^{\otimes k}}(d_1 \gamma d_2^T \gamma^{-1}),
\end{equation}
or equivalently, the RHS of equation \eqref{eq: Two Point Function of Partition Algebra Observables} for the two-point function.

It is useful to consider the simpler case
\begin{equation}
	\Tr_{V_N^{\otimes k}}(d_1 d_2^T).
\end{equation}
This trace can be computed in terms of the number of connected components in the diagram $d_1 \merge d_2$, given by a diagram with all the edges of $d_1$ and $d_2$. In the mathematics literature, this operation is called the join on the partition lattice (see \cite{birkhoff1940lattice}).
It is given by 
\begin{equation}
	\Tr_{V_N^{\otimes k}}(d_1 d_2^T) = N^{c(d_1 \merge d_2)}. \label{eq: Trace counts components of join}
\end{equation}
where $c(d)$ is the number of connected components in the diagram $d$.
Examples of the join operation are
\begin{equation}
	\PAdiagram{2}{1/-1} \merge \PAdiagram{2}{1/-2} = \PAdiagram{2}{-1/1,1/-2}, \qq{and} \PAdiagram{2}{1/-1} \merge \PAdiagram{2}{1/2} = \PAdiagram{2}{1/-1,1/2}.
\end{equation}
Examples of $c(d)$ are
\begin{equation}
	c\qty(\PAdiagram{2}{1/-1, 2/-2}) = 2, \quad c\qty(\PAdiagram{2}{1/-1}) = 3.
\end{equation}
To illustrate equation \eqref{eq: Trace counts components of join} consider the following pair of diagrams
\begin{equation}
	\qty(\PAdiagram{2}{1/-1})^{i_{1'}i_{2'}}_{i_1 i_2} = \delta_{i_1}^{i_{1'}}, \quad \qty(\PAdiagram{2}{2/-2})^{i_{1'}i_{2'}}_{i_1 i_2} = \delta_{i_2}^{i_{2'}}. 
\end{equation}
The join is given by
\begin{equation}
\quad \qty(\PAdiagram{2}{1/-1} \merge \PAdiagram{2}{-2/2})^{i_{1'}i_{2'}}_{i_1 i_2} = \qty(\PAdiagram{2}{1/-1, 2/-2})^{i_{1'}i_{2'}}_{i_1 i_2} = \delta_{i_1}^{i_{1'}} \delta_{i_2}^{i_{2'}}.
\end{equation}
The diagram multiplication gives
\begin{equation}
	\Tr_{V_N^{\otimes 2}}\qty(\PAdiagram{2}{1/-1} \qty(\PAdiagram{2}{-2/2})^T) = \Tr_{V_N^{\otimes 2}}\qty(\begin{aligned}
		&\PAdiagram{2}{1/-1} \\
		&\PAdiagram{2}{-2/2}
	\end{aligned}) = \Tr_{V_N^{\otimes 2}}\qty(\PAdiagram{2}{}) = N^2,
\end{equation}
while the corresponding expression using the join gives
\begin{equation}
	\Tr_{V_N^{\otimes 2}}\qty(\PAdiagram{2}{1/-1} \qty(\PAdiagram{2}{-2/2})^T)  = N^{c\qty(\mbox{$\PAdiagram{2}{1/-1} \merge \PAdiagram{2}{-2/2}$})}= N^{c\qty(\PAdiagram{2}{1/-1, 2/-2})} =N^2.
\end{equation}

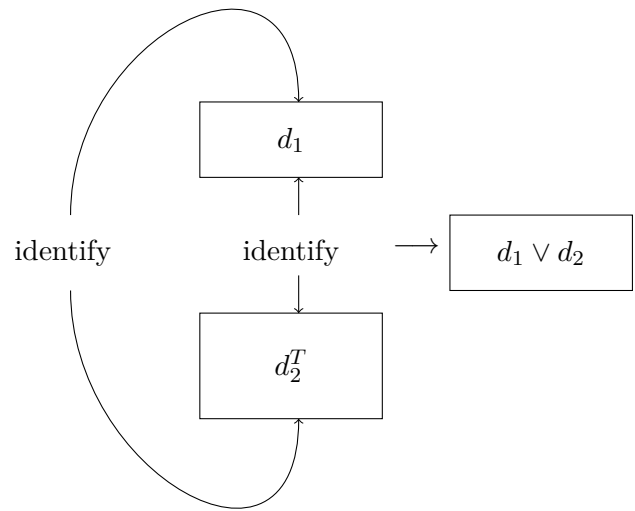
\begin{figure}
	\centering
	\begin{tikzpicture}[baseline=(base), yscale=2] 
		\coordinate (size) at (2,0);
		\coordinate (sep) at (0,0);
		\coordinate (mar) at (0.2,0);
		\coordinate (l1) at (0,1) {}; \coordinate (l2) at ($(l1)+(size)$) {}; \coordinate (l3) at ($(l2)+(size)+(sep)$) {}; \coordinate (l4) at ($(l3)+(size)$) {};
		\coordinate (m1) at (0,1.5) {}; \coordinate (m2) at ($(m1)+(size)$) {}; \coordinate (m3) at ($(m2)+(size)+(sep)$) {}; \coordinate (m4) at ($(m3)+(size)$) {};
		\coordinate (n1) at (0,1.75) {}; \coordinate (n2) at ($(n1)+(size)$) {}; \coordinate (n3) at ($(n2)+(size)+(sep)$) {}; \coordinate (n4) at ($(n3)+(size)$) {};
		\coordinate (o1) at (0,2.25) {}; \coordinate (o2) at ($(o1)+(size)$) {}; \coordinate (o3) at ($(o2)+(size)+(sep)$) {}; \coordinate (o4) at ($(o3)+(size)$) {};
		\coordinate (p1) at (0,2.5) {}; \coordinate (p2) at ($(p1)+(size)$) {}; \coordinate (p3) at ($(p2)+(size)+(sep)$) {}; \coordinate (p4) at ($(p3)+(size)$) {};
		\coordinate (q1) at (0,3) {}; \coordinate (q2) at ($(q1)+(size)$) {}; \coordinate (q3) at ($(q2)+(size)+(sep)$) {}; \coordinate (q4) at ($(q3)+(size)$) {};
		\coordinate (r1) at (0,3.25) {}; \coordinate (r2) at ($(r1)+(size)$) {}; \coordinate (r3) at ($(r2)+(size)+(sep)$) {}; \coordinate (r4) at ($(r3)+(size)$) {};
		\fill[fill=white] ($(o1)-(mar)$) rectangle node{identify} ($(n2)+(mar)$);
		\fill[fill=white] ($(o1)-(mar)-(3,0)$) rectangle node{identify} ($(n2)+(mar)-(3,0)$);
		\draw[<-] ($(q1)+(1,0)+(0.1,0)$) to[out=90,in=90] ($(o1)+(0.1,0)-(2,0)$);
		\draw[<-] ($(l1)+(1,0)+(0.1,-0.1)$) to[out=-90,in=-90] ($(n1)+(0.1,0)-(2,0)$);
		\draw[fill=white] ($(q1)-(mar)$) rectangle node{$d_1$} ($(p2)+(mar)$); \draw[fill=white] ($(m1)-(mar)+(0,0.1)$) rectangle node{$d_2^T$} ($(l2)+(mar)-(0,0.1)$);
		\draw[->] 	($(o1)+(1,0)+(0.1,0)$) -- ($(p1)+(1,0)+(0.1,0)$);
		\draw[->] 	($(n1)+(1,0)+(0.1,0.1)$) -- ($(m1)+(1,0)+(0.1,0.1)$);
		\coordinate (base) at ($(o1)-(0,0.25)$);
	\end{tikzpicture}
	$\longrightarrow$
	\begin{tikzpicture}[baseline=(base), yscale=2]
		\coordinate (size) at (2,0);
		\coordinate (sep) at (0,0);
		\coordinate (mar) at (0.2,0);
		\coordinate (l1) at (0,1) {}; \coordinate (l2) at ($(l1)+(size)$) {}; \coordinate (l3) at ($(l2)+(size)+(sep)$) {}; \coordinate (l4) at ($(l3)+(size)$) {};
		\coordinate (m1) at (0,1.5) {}; \coordinate (m2) at ($(m1)+(size)$) {}; \coordinate (m3) at ($(m2)+(size)+(sep)$) {}; \coordinate (m4) at ($(m3)+(size)$) {};
		\coordinate (n1) at (0,1.75) {}; \coordinate (n2) at ($(n1)+(size)$) {}; \coordinate (n3) at ($(n2)+(size)+(sep)$) {}; \coordinate (n4) at ($(n3)+(size)$) {};
		\coordinate (o1) at (0,2.25) {}; \coordinate (o2) at ($(o1)+(size)$) {}; \coordinate (o3) at ($(o2)+(size)+(sep)$) {}; \coordinate (o4) at ($(o3)+(size)$) {};
		\coordinate (p1) at (0,2.5) {}; \coordinate (p2) at ($(p1)+(size)$) {}; \coordinate (p3) at ($(p2)+(size)+(sep)$) {}; \coordinate (p4) at ($(p3)+(size)$) {};
		\coordinate (q1) at (0,3) {}; \coordinate (q2) at ($(q1)+(size)$) {}; \coordinate (q3) at ($(q2)+(size)+(sep)$) {}; \coordinate (q4) at ($(q3)+(size)$) {};
		\coordinate (r1) at (0,3.25) {}; \coordinate (r2) at ($(r1)+(size)$) {}; \coordinate (r3) at ($(r2)+(size)+(sep)$) {}; \coordinate (r4) at ($(r3)+(size)$) {};
		\draw[fill=white] ($(o1)-(mar)$) rectangle node{$d_1 \vee d_2$} ($(n2)+(mar)$);
		\coordinate (base) at ($(o1)-(0,0.25)$); 
	\end{tikzpicture}
	\caption{By identifying the bottom vertices of $d_2^T$ with the top vertices of $d_1$, and the top vertices of $d_2^T$ with the bottom vertices of $d_1$, we have constructed a diagram with all the edges of $d_1$ together with all the edges of $d_2$. This is equal to the diagram $d_1 \merge d_2$. This Figure is a central Koan of the factorization proof. }
	\label{fig: join from trace}
\end{figure}

To prove this, recall that every edge in a diagram corresponds to a Kronecker delta when acting on $V_N^{\otimes k}$ (see examples in \eqref{eq: examples of diagram tensors}). Consequently
\begin{equation}
	(d_1 \merge d_2)_{i_1 \dots i_k}^{i_{1'}\dots i_{k'}} = (d_1)_{i_1 \dots i_k}^{i_{1'}\dots i_{k'}}(d_2)_{i_1 \dots i_k}^{i_{1'}\dots i_{k'}}.
\end{equation}
It follows that
\begin{equation}
	\begin{aligned}
		\Tr_{V_N^{\otimes k}}(d_1 d_2^T) &= \sum_{\substack{i_1, \dots, i_k \\ i_{1'}, \dots i_{k'}}} (d_1)_{i_1 \dots i_k}^{i_{1'} \dots i_{k'}}(d_2^T)_{i_{1'} \dots i_{k'}}^{i_{1} \dots i_{k}} = \sum_{\substack{i_1, \dots, i_k \\ i_{1'}, \dots i_{k'}}} (d_1)_{i_1 \dots i_k}^{i_{1'} \dots i_{k'}}(d_2)^{i_{1'} \dots i_{k'}}_{i_{1} \dots i_{k}} \\
		&= \sum_{\substack{i_1, \dots, i_k \\ i_{1'}, \dots i_{k'}}} (d_1 \merge d_2)^{i_{1'} \dots i_{k'}}_{i_{1} \dots i_{k}}.
	\end{aligned}
\end{equation}
Equivalently, the diagrammatic representation of a trace identifies the bottom vertices with the top vertices,
\begin{equation}
	\Tr_{V_N^{\otimes k}}(d_1 d_2^T)  = \begin{tikzpicture}[baseline = (m1)]
		\coordinate (base) at (0,-.25); 
		\coordinate (size) at (2,0);
		\coordinate (sep) at (0,0);
		\coordinate (mar) at (0.2,0);
		\coordinate (i1) at (0,0) {}; \coordinate (i2) at ($(i1)+(size)$) {}; \coordinate (i3) at ($(i2)+(size)+(sep)$) {}; \coordinate (i4) at ($(i3)+(size)$) {};
		\coordinate (j1) at (0,.25) {}; \coordinate (j2) at ($(j1)+(size)$) {}; \coordinate (j3) at ($(j2)+(size)+(sep)$) {}; \coordinate (j4) at ($(j3)+(size)$) {};
		\coordinate (k1) at (0,.75) {}; \coordinate (k2) at ($(k1)+(size)$) {}; \coordinate (k3) at ($(k2)+(size)+(sep)$) {}; \coordinate (k4) at ($(k3)+(size)$) {};
		\coordinate (l1) at (0,1) {}; \coordinate (l2) at ($(l1)+(size)$) {}; \coordinate (l3) at ($(l2)+(size)+(sep)$) {}; \coordinate (l4) at ($(l3)+(size)$) {};
		\coordinate (m1) at (0,1.5) {}; \coordinate (m2) at ($(m1)+(size)$) {}; \coordinate (m3) at ($(m2)+(size)+(sep)$) {}; \coordinate (m4) at ($(m3)+(size)$) {};
		\coordinate (n1) at (0,1.75) {}; \coordinate (n2) at ($(n1)+(size)$) {}; \coordinate (n3) at ($(n2)+(size)+(sep)$) {}; \coordinate (n4) at ($(n3)+(size)$) {};
		\coordinate (o1) at (0,2.25) {}; \coordinate (o2) at ($(o1)+(size)$) {}; \coordinate (o3) at ($(o2)+(size)+(sep)$) {}; \coordinate (o4) at ($(o3)+(size)$) {};
		\coordinate (p1) at (0,2.5) {}; \coordinate (p2) at ($(p1)+(size)$) {}; \coordinate (p3) at ($(p2)+(size)+(sep)$) {}; \coordinate (p4) at ($(p3)+(size)$) {};
		\coordinate (q1) at (0,3) {}; \coordinate (q2) at ($(q1)+(size)$) {}; \coordinate (q3) at ($(q2)+(size)+(sep)$) {}; \coordinate (q4) at ($(q3)+(size)$) {};
		\coordinate (r1) at (0,3.25) {}; \coordinate (r2) at ($(r1)+(size)$) {}; \coordinate (r3) at ($(r2)+(size)+(sep)$) {}; \coordinate (r4) at ($(r3)+(size)$) {};
		\draw (k1) -- (p1); \draw (k2) -- (p2); 
		\draw[fill=white] ($(o1)-(mar)$) rectangle node{$d_1$} ($(n2)+(mar)$);
		\draw[fill=white] ($(m1)-(mar)+(0,0.1)$) rectangle node{$d_2^T$} ($(l2)+(mar)-(0,0.1)$);	
		\draw ($(k1)-(mar)$) -- ($(k2)+(mar)$);	 \draw ($(p1)-(mar)$) -- ($(p2)+(mar)$);
	\end{tikzpicture}.
\end{equation}
Taken literally, this means that we identify the bottom vertices of $d_2^T$ with the top vertices of $d_1$, and the top vertices of $d_2^T$ with the bottom vertices of $d_1$. The diagram constructed in this manner has all the edges of $d_1$ together with all the edges of $d_2$, which is precisely equal to $d_1 \merge d_2$. See figure \ref{fig: join from trace} for an illustration.

To complete the proof we show that
\begin{equation}
	\Tr_{V_N^{\otimes k}}(d_1 d_2^T)  = \begin{tikzpicture}[baseline = (m1)]
		\coordinate (base) at (0,-.25); 
		\coordinate (size) at (2,0);
		\coordinate (sep) at (0,0);
		\coordinate (mar) at (0.2,0);
		\coordinate (i1) at (0,0) {}; \coordinate (i2) at ($(i1)+(size)$) {}; \coordinate (i3) at ($(i2)+(size)+(sep)$) {}; \coordinate (i4) at ($(i3)+(size)$) {};
		\coordinate (j1) at (0,.25) {}; \coordinate (j2) at ($(j1)+(size)$) {}; \coordinate (j3) at ($(j2)+(size)+(sep)$) {}; \coordinate (j4) at ($(j3)+(size)$) {};
		\coordinate (k1) at (0,.75) {}; \coordinate (k2) at ($(k1)+(size)$) {}; \coordinate (k3) at ($(k2)+(size)+(sep)$) {}; \coordinate (k4) at ($(k3)+(size)$) {};
		\coordinate (l1) at (0,1) {}; \coordinate (l2) at ($(l1)+(size)$) {}; \coordinate (l3) at ($(l2)+(size)+(sep)$) {}; \coordinate (l4) at ($(l3)+(size)$) {};
		\coordinate (m1) at (0,1.5) {}; \coordinate (m2) at ($(m1)+(size)$) {}; \coordinate (m3) at ($(m2)+(size)+(sep)$) {}; \coordinate (m4) at ($(m3)+(size)$) {};
		\coordinate (n1) at (0,1.75) {}; \coordinate (n2) at ($(n1)+(size)$) {}; \coordinate (n3) at ($(n2)+(size)+(sep)$) {}; \coordinate (n4) at ($(n3)+(size)$) {};
		\coordinate (o1) at (0,2.25) {}; \coordinate (o2) at ($(o1)+(size)$) {}; \coordinate (o3) at ($(o2)+(size)+(sep)$) {}; \coordinate (o4) at ($(o3)+(size)$) {};
		\coordinate (p1) at (0,2.5) {}; \coordinate (p2) at ($(p1)+(size)$) {}; \coordinate (p3) at ($(p2)+(size)+(sep)$) {}; \coordinate (p4) at ($(p3)+(size)$) {};
		\coordinate (q1) at (0,3) {}; \coordinate (q2) at ($(q1)+(size)$) {}; \coordinate (q3) at ($(q2)+(size)+(sep)$) {}; \coordinate (q4) at ($(q3)+(size)$) {};
		\coordinate (r1) at (0,3.25) {}; \coordinate (r2) at ($(r1)+(size)$) {}; \coordinate (r3) at ($(r2)+(size)+(sep)$) {}; \coordinate (r4) at ($(r3)+(size)$) {};
		\draw (k1) -- (p1); \draw (k2) -- (p2); 
		\draw[fill=white] ($(o1)-(mar)$) rectangle node{$d_1$} ($(n2)+(mar)$);
		\draw[fill=white] ($(m1)-(mar)+(0,0.1)$) rectangle node{$d_2^T$} ($(l2)+(mar)-(0,0.1)$);	
		\draw ($(k1)-(mar)$) -- ($(k2)+(mar)$);	 \draw ($(p1)-(mar)$) -- ($(p2)+(mar)$);
	\end{tikzpicture} = \sum_{\substack{i_1, \dots, i_k \\ i_{1'}, \dots i_{k'}}} (d_1 \merge d_2)^{i_{1'} \dots i_{k'}}_{i_{1} \dots i_{k}} = N^{c(d_1 \merge d_2)}.
\end{equation}
Let $b_1, \dots, b_l$ be sets containing the vertices of connected components of $d_1 \merge d_2$. Then,
\begin{equation}
	\sum_{\substack{i_1, \dots, i_k \\ i_{1'}, \dots i_{k'}}} (d_1 \merge d_2)^{i_{1'} \dots i_{k'}}_{i_{1} \dots i_{k}} = \qty(\sum_{b_1} 1) \qty(\sum_{b_2} 1) \dots \qty(\sum_{b_l} 1) = N^{c(d_1 \merge d_2)},
\end{equation}
where the sums over connected components correspond to sums where the indices in each component are set equal. For example, if $b_1 = \{1,3,5',8\}$ then
\begin{equation}
	\sum_{b_1} 1 \equiv \sum_{i_1,i_3,i_{5'},i_8} \delta_{i_1 i_3}\delta_{i_3 i_{5'}}\delta_{i_{5'}i_8} = \sum_{i_1 = i_3 = i_{5'}=i_8} 1 = N.
\end{equation}

\subsection{Factorization for trace form on $P_k(N)$}
\label{sec:facPa}

The proof of the following version of factorization
\begin{equation}
	\frac{\Tr_{V_N^{\otimes k}}(d_1 d_2^T)}{\sqrt{\Tr_{V_N^{\otimes k}}(d_1 d_1^T) \Tr_{V_N^{\otimes k}}(d_2 d_2^T)}} = \begin{cases}
		1 + O(1/\sqrt{N}) \qq{if $d_1 = d_2$,}\\
		0 + O(1/\sqrt{N}) \qq{if $d_1 \neq d_2$,}
	\end{cases} \label{eq: simple factorization case}
\end{equation} 
contains most of the essential ingredients necessary for the 1-matrix case. This is a useful warm-up exercise and, as we will see in section \ref{sec:factr}, a special case of factorization in multi-matrix models. This equation \eqref{eq: simple factorization case} is related to the properties of the distance function defined in proposition 3.1 of \cite{gabrielcombinatorial1}.\footnote{We thank Franck Gabriel for this observation.}

The factorization in equation \eqref{eq: simple factorization case}  is a consequence of the following 
\begin{equation}
	\begin{aligned}
		&2c(d_1 \merge d_2) = c(d_1 \merge d_1) + c(d_2 \merge d_2) = c(d_1) + c(d_2) \qq{if} d_1 = d_2,\\
		&2c(d_1 \merge d_2) < c(d_1 \merge d_1) + c(d_2 \merge d_2) = c(d_1) + c(d_2) \qq{if} d_1 \neq d_2,		
	\end{aligned} \label{eq: non-symmetric inequality}
\end{equation}
where we have used {${c(d_1 \merge d_1) + c(d_2 \merge d_2) = c(d_1) + c(d_2)}$} since $d \merge d = d$.
We will prove \eqref{eq: non-symmetric inequality} by separating the general pairs $d_1 , d_2$  into three distinct cases:
\begin{enumerate}
	\item If $d_1$ only contains edges that are also contained in $d_2$, but $d_1 \neq d_2$, we write $d_1 < d_2$. For example,
	\begin{equation}
		\PAdiagram{1}{} < \PAdiagram{1}{1/-1}, \qq{and} \PAdiagram{2}{1/2} < \PAdiagram{2}{-1/-2,1/2}.
	\end{equation}
	In this case, $d_1 \merge d_2 = d_2$ and it follows that,
	\begin{equation}\label{d1ind2conn} 
		c(d_1 \merge d_2) = c(d_2).
	\end{equation}
	Note that $d_1 < d_2$ implies $c(d_1) > c(d_2)$. Therefore,
	\begin{equation}
		2c(d_1 \merge d_2) = c(d_2) + c(d_2) < c(d_1) + c(d_2).
	\end{equation}
	Since the LHS and RHS are symmetric under exchanging $d_1 \leftrightarrow d_2$, the inequality {${2c(d_1 \merge d_2) <  c(d_1) + c(d_2)}$} holds for $d_2 < d_1$ as well.
	\item Suppose $d_1 \neq d_2$ and  that there is no set of edges that can be added to $d_1$ to turn it into $d_2$, nor is there a set of edges that can be added to $d_2$ to obtain $d_1$. Then, we say that $d_1$ and $d_2$ are incomparable. We denote this by $d_1 \not\lesseqqgtr d_2$. The following diagrams are examples of incomparable diagrams
	\begin{equation}
		\PAdiagram{2}{-1/-2,1/2} \not\lesseqqgtr \PAdiagram{2}{1/-2,2/-1},  \qq{and} \PAdiagram{2}{-1/-2}  \not\lesseqqgtr \PAdiagram{2}{1/2,2/-2}. \label{eq: incomparable diagrams}
	\end{equation}
 In this incomparable case, we have 
	\bea\label{incompconn}  
	c ( d_1 \merge d_2 )  < c ( d_1 ) \hbox{ and }   	c ( d_1 \merge d_2 ) < c ( d_2 ) 
	\eea
	since the forming of  the join involves adding to  $d_1 $, additional edges creating connections which did not exist in  $d_1$, or  alternatively adding to $d_2$ additional edges that did not exist in $d_2$.  
	Consequently we have the inequality 
	\begin{equation}\label{IncompIneq} 
		2c(d_1 \merge d_2) < c(d_1) + c(d_2).
	\end{equation}
	\item If $d_1 = d_2$ we have
	\begin{equation}
		c(d_1 \merge d_2) = c(d_1 \merge d_1) = c(d_1) = c(d_2),
	\end{equation}
	and therefore,
	\begin{equation}
		2c(d_1 \merge d_2)  = c(d_1) + c(d_2).
	\end{equation}
\end{enumerate}
To summarize, $2c(d_1 \merge d_2) \leq c(d_1) + c(d_2)$ with equality if and only if $d_1 = d_2$.

As a corollary of the above discussion, which will be useful in the next sub-section, note that 
if we consider a fixed diagram $d_1$ and  a family of diagrams $d_3$ with fixed $c( d_3) $ such that   $ c(d_1 ) > c (d_3) $, then we have for each $d_3$ in the family one of the following  
\bea\label{Corr1}
&& c ( d_1 \merge d_3 ) <  c ( d_3 )~~~  \hbox{  if  } d_1 \not\lesseqqgtr  d_3 \cr 
&& c ( d_1 \merge d_3 ) = c ( d_3)  ~~~  \hbox{ if } d_1 < d_3 
\eea
 This follows from \eqref{d1ind2conn} and \eqref{incompconn}. 

\subsection{Factorization for PIMOs}
\label{sec:factr} 
The 1-matrix connected two-point function \eqref{eq: Two Point Function of Partition Algebra Observables} includes a sum over $\gamma \in S_k$,
\begin{equation}
	\expval{\normord{\mathcal{\hat{O}}_{d_1}}\normord{\mathcal{\hat{O}}_{d_2}}} = \frac{\sum_{\gamma_1 \in S_k} N^{c(d_1 \merge \gamma_1 d_2 \gamma_1^{-1})}}{\sqrt{\sum_{\gamma_2 \in S_k} N^{c(d_1 \merge \gamma_2 d_1 \gamma_2^{-1})}\sum_{\gamma_3 \in S_k} N^{c(d_2 \merge \gamma_3 d_2 \gamma_3^{-1})}}}.
\end{equation}
Large $N$ factorization of PIMOs follows from the inequalities
\begin{equation}
	\begin{aligned}
		2\max_{\gamma_1} c(d_1 \merge \gamma_1 d_2  \gamma_1^{-1}) = \max_{\gamma_2} c(d_1 \merge \gamma_2 d_1  \gamma_2^{-1}) + \max_{\gamma_3}  c(d_2 \merge \gamma_3 d_2  \gamma_3^{-1}) \qq{if} [d_1] = [d_2],\\
		2\max_{\gamma_1} c(d_1 \merge \gamma_1 d_2  \gamma_1^{-1}) < \max_{\gamma_2} c(d_1 \merge \gamma_2 d_1  \gamma_2^{-1}) + \max_{\gamma_3}  c(d_2 \merge \gamma_3 d_2  \gamma_3^{-1}) \qq{if} [d_1] \neq [d_2]
	\end{aligned} \label{eq: Large N factorisation inequality}
\end{equation}

The first step in proving equation \eqref{eq: Large N factorisation inequality} is to simplify the terms on the RHS.
The inequalities in equation \eqref{eq: non-symmetric inequality} imply that $c(d \merge \gamma d  \gamma^{-1})$ is maximized when $d = \gamma d \gamma^{-1}$. Of course, the identity permutation always satisfies this equality. Therefore,
\begin{equation}
	\max_{\gamma} c(d \merge \gamma d  \gamma^{-1}) = c(d).
\end{equation}
We are left with proving
\begin{equation}
	\begin{aligned}
		2\max_{\gamma } c(d_1 \merge \gamma  d_2  \gamma^{-1}) = c(d_1) + c(d_2) \qq{if} [d_1] = [d_2],\\
		2\max_{\gamma } c(d_1 \merge \gamma  d_2  \gamma^{-1}) < c(d_1) + c(d_2) \qq{if} [d_1] \neq [d_2]
	\end{aligned} \label{eq: Large N factorisation inequality 2}
\end{equation}
We employ the same strategy as before, and consider the three distinct cases.
\begin{enumerate}
	\item Suppose $c(d_1) > c(d_2)$, and consider the diagrams $ \gamma d_2 \gamma^{-1} $ for $ \gamma \in S_k$. We have $ c ( d_1 ) > c ( \gamma d_2 \gamma^{-1} )  = c ( d_2 ) $. 
Assume $d_1 , d_2 $ are such there exists some $ \gamma^* $ such that 	$d_1 < \gamma^* d_2 (\gamma^*)^{-1}$. For any such $ \gamma^*$, the equality in \eqref{Corr1}
	implies that 
	\bea 
	2 c(d_1 \merge \gamma^* d_2 (\gamma^*)^{-1}) = 2c(d_2) < c(d_1) + c(d_2).
	\eea 
Any $\gamma $ not satisfying this condition leads to 
$ d_1 \not\lesseqqgtr \gamma d_2 \gamma^{-1} $, and the inequality in \eqref{Corr1} implies that 
\bea 
2 c(d_1 \merge \gamma d_2 \gamma^{-1}) <  2c(d_2)
\eea
This implies that
	\begin{equation}
		2\max_{\gamma} c(d_1 \merge \gamma d_2 \gamma^{-1}) = 2 c(d_1 \merge \gamma^* d_2 (\gamma^*)^{-1}) = 2c(d_2) < c(d_1) + c(d_2). 	\label{eq: tau max case 1}
	\end{equation}
	The pair
	\begin{equation}
		d_1 = \PAdiagram{2}{-1/1}, \quad d_2 = \PAdiagram{2}{1/2,-2/2},
	\end{equation}
	is an example of this case since
	\begin{equation}
		\PAdiagram{2}{-1/1} < \PAdiagram{2}{1/2,-1/1} = \begin{aligned}
			&\PAdiagram{2}{-1/2,-2/1} \\
			&\PAdiagram{2}{1/2,-2/2}\\
			&\PAdiagram{2}{-1/2,-2/1}
		\end{aligned}.
	\end{equation}
	 The argument is identical for the case where $c(d_1) < c(d_2)$, and there  exists some 
	  $\gamma^*  \in S_k$ such that $d_2 < \gamma^* d_1 (\gamma^*)^{-1}$. In this case, by renaming $ d_1 \leftrightarrow d_2 $ in \eqref{eq: tau max case 1}, we have 
	  \bea 
	  	2\max_{\gamma} c(d_2 \merge \gamma d_1 \gamma^{-1}) = 2 c(d_2 \merge \gamma^* d_1 (\gamma^*)^{-1}) = 2c(d_1) < c(d_1) + c(d_2). 
	  \eea
	  Using the symmetry of the inner product \eqref{Symmetry} it follows 
\bea 
2\max_{\gamma} c(d_1 \merge \gamma d_2 \gamma^{-1}) < c(d_1) + c(d_2). 
\eea	  
	\item Secondly, consider the case of incomparability,
	\begin{equation}
		d_1 \not\lesseqqgtr \gamma d_2 \gamma^{-1} \quad \forall \gamma \in S_k.
	\end{equation}
	Recall that for incomparable diagrams \eqref{IncompIneq},  
	\begin{equation}
		2c(d_1 \merge \gamma d_2 \gamma^{-1}) < c(d_1) + c(\gamma d_2 \gamma^{-1}) = c(d_1)+ c(d_2),
	\end{equation}
	where the last equality follows because conjugation by a permutation does not change the number of connected components.
	Therefore
	\begin{equation}
		2\max_{\gamma} c(d_1 \merge \gamma d_2 \gamma^{-1}) < c(d_1) + c(d_2),		
	\end{equation}
	in this case as well.
	\item When $d_1 = \gamma d_2 \gamma^{-1}$ for some $\gamma \in S_k$, the bound is saturated and \begin{equation}
		2\max_{\gamma} c(d_1 \merge \gamma d_2 \gamma^{-1}) = 2c(d_1).
	\end{equation}
\end{enumerate}
The condition $d_1 = \gamma d_2 \gamma^{-1}$ implies $[d_1] = [d_2]$.	We have proven the inequalities in equation \eqref{eq: Large N factorisation inequality}. As a consequence, we have large $N$ factorization of permutation invariant matrix observables.

\subsection{Factorization for multi-matrix observables}
The above argument generalizes to multi-matrix models. Let $M^{(f)}$ be $n$ matrices with flavour label $f=1,\dots,n$ and second moment
\begin{equation}
	\expval{(M^{(f)})^{i}{}_j(M^{(f')})^{k}{}_l}_{\con} = \delta^{ff'}\delta^{ik}\delta_{jl}.
\end{equation}
Permutation invariant multi-matrix observables of degree $k=k_1 + k_2 +\dots + k_n$, where $k_f$ is the degree of matrix $M^{(f)}$, are constructed using partition algebra elements. Multi-matrix observables are labelled by $\vec{k} = (k_1, \dots, k_n)$ and $d \in P_k(N)$
\begin{equation}
	\mathcal{O}_{\vec{k},d} = \Tr_{V_N^{\otimes k}}( ( M^{(1)} )^{\otimes k_1} \dots ( M^{(n)} )^{\otimes k_n} d).
\end{equation}
As before, we have bosonic symmetry.
For any $\gamma \in S_{\vec{k}} \equiv S_{k_1} \times \dots \times S_{k_n}$ observables are invariant
\begin{equation}
	\begin{aligned}
		\mathcal{O}_{\vec{k},\gamma d \gamma^{-1}} &= \Tr_{V_N^{\otimes k}}( ( M^{(1)} )^{k_1} \otimes \dots \otimes ( M^{(n)} )^{k_n} \gamma d \gamma^{-1}) \\
		& = \Tr_{V_N^{\otimes k}}(\gamma^{-1} ( M^{(1)} )^{k_1} \otimes \dots\otimes  ( M^{(n)} )^{k_n} \gamma d ) \\
		&=\Tr_{V_N^{\otimes k}}( ( M^{(1)} )^{k_1} \otimes \dots \otimes ( M^{(n)} )^{k_n} d) = \mathcal{O}_{\vec{k},d}.
	\end{aligned}
\end{equation}
Multi-matrix observables are in one-to-one correspondence with partition algebra equivalence classes
\begin{equation}
	[d] = \{\gamma d \gamma^{-1} \, \vert \, \gamma \in S_{\vec{k}}\}.
\end{equation}

Wick contractions vanish unless the flavour indices match, and the sum over $\gamma \in S_k$ reduces to a sum over $\gamma \in S_{\vec{k}}$
\begin{equation}
	\expval{\mathcal{O}_{\vec{k},d_1}\mathcal{O}_{\vec{k}',d_2}}_{\con} = \delta_{\vec{k} \vec{k'}} \sum_{\gamma \in S_{\vec{k}}} \Tr_{V_N^{\otimes k}}( d_1 \gamma d_2 \gamma^{-1}) =\delta_{\vec{k} \vec{k'}} \sum_{\gamma \in S_{\vec{k}}} N^{c(d_1 \merge \gamma d_2 \gamma^{-1})}.
\end{equation}
The same argument holds for the inequality
\begin{equation}
	2\max_{\gamma} c(d_1 \merge \gamma d_2 \gamma^{-1}) \leq \max_{\gamma} c(d_1 \merge \gamma d_2 \gamma^{-1}) + \max_{\gamma} c(d_1 \merge \gamma d_2 \gamma^{-1}) = c(d_1) + c(d_2).
\end{equation}
It is saturated if and only if there exists a $\gamma \in S_{\vec{k}}$ such that $d_1 = \gamma d_2 \gamma^{-1}$. That is, if and only if $[d_1] = [d_2]$ or
\begin{equation}
	\mathcal{O}_{\vec{k},d_1} = \mathcal{O}_{\vec{k},d_2}.
\end{equation}
To summarize we have
\begin{equation}
	\expval{\hat{\mathcal{O}}_{\vec{k},d_1} \hat{\mathcal{O}}_{\vec{k}',d_2}}_{\con} = \delta_{\vec{k}\vec{k}'} \times \begin{cases}
		1 + O(1/\sqrt{N}) \qq{if $[d_1] = [d_2]$,} \\
		0 + O(1/\sqrt{N}) \qq{if $[d_1] \neq [d_2]$,}
	\end{cases}
\end{equation}
for permutation invariant multi-matrix observables in the above Gaussian $O(N)$ model.

Note that in the case $n = k$, $k_f = 1$ (all matrices distinct), we have
\begin{equation}
	S_{\vec{k}} = \underbrace{S_1 \times \dots \times S_1}_{n}.
\end{equation}
Therefore, the sum over Wick contractions reduces to a single element (the identity element).
The corresponding two-point function is the first case we considered (equation \eqref{eq: simple factorization case}).

Finally, we observe that the proof of factorization presented here for general observables labelled by partition algebra diagrams specializes to  a new way of thinking about factorization 
 in the case of matrix invariants with continuous symmetry, where the partition algebra diagrams specialize to permutations. The previously known 
 proof based on permutation products can be understood, in the one-matrix case,  
from the equation 
\bea 
\langle \cO_{\sigma_1 } ( Z ) \cO_{ \sigma_2} ( Z^{ \dagger} ) \rangle 
= {  k! \over |T_1| |T_2| }  \sum_{ \sigma_1' \in T_1 } \sum_{ \sigma_2' \in T_2 } \sum_{ \sigma_3 \in S_k} 
\delta ( \sigma_1' \sigma_2' \sigma_3 ) N^{ C_{ \sigma_3} }  
\eea
This equation is derived and explained as eqn. (2.12) in \cite{Ramgoolam2016} (multi-matrix generalisations are discussed in references therein).   Gauge invariant operators are labelled by permutations $ \sigma_1, \sigma_2$ in conjugacy classes $T_1 , T_2$, while $|T_1| , |T_2| $ are the sizes of these conjugacy classes. Large $N$ factorisation follows from the fact that the largest power of $N$ comes from the case where $ \sigma_3$ is the identity and this only occurs when $ T_1 = T_2$. In the present way of looking at permutations as special cases of partition algebra diagrams, permutations belonging to distinct conjugacy classes are always incomparable in  the partial order on set partitions associated to the diagrams.  This corresponds to Case 2 in of the proofs in sections \ref{sec:facPa} and \ref{sec:factr}. 

\section{Discussion}

In this paper we considered $S_N$ invariant matrix models. These can be viewed as generalisations of their more familiar  cousins invariant under continuous symmetries. The most general $S_N$ invariant Gaussian matrix model is specified by a 13-dimensional parameter space. We have shown that there exists a four-dimensional subspace of the 13-dimensional  parameter space in which the $S_N$ symmetry is enhanced to $O(N)$. The parameter limit in which this enhancement takes place is given by equation \eqref{eqn:5in13}. The special case of the simplest $O(N)$ invariant Gaussian \eqref{eqn:simplestON} arises at the parameters given in \eqref{eqn:1in13}. 

The factorisation property of multi-trace matrix observables invariant under continuous symmetries such as $U(N)$  in the large $N$ limit is a well known result.  We  have shown that this continues to hold for $S_N$ invariant observables. In the $U(N)$ case this can be seen using properties of the symmetric group by exploiting the Schur-Weyl duality of $U(N)$ and $S_k$ in order to establish a correspondence between observables and conjugacy classes of $S_k$. Analogously, we gave a description of the permutation invariant matrix polynomial functions in terms of a diagram basis for partition algebras. We used the inner product on the permutation invariant polynomials arising from the simplest $O(N)$ invariant action, and proved large $N$ factorisation. The partial order on the diagram basis elements, which can itself be  described by a Hasse diagram, plays a role in the proof of factorisation.

As explained in the introduction, there are two 
guiding principles in this paper: the analogies between results for $U(N)$ invariant matrix models and $S_N$ invariant models, and the Schur-Weyl dual algebras of these respective invariances. These principles can be exploited in a number of natural generalisations of the results in this paper. For example, they are applicable to 
one-dimensional quantum mechanics models of matrices  (see our companion paper \cite{MQMPartAlg}). They are also applicable to tensor models: this is being developed in  \cite{PermTensor}. Permutation invariant random matrix distributions have been considered using techniques from probability theory \cite{Malle}. It would be interesting to investigate the implications of the factorisation results presented here in that context.

The $1/N$ expansion of simple correlators in  $U(N)$ invariant matrix models has a geometrical interpretation in terms of Belyi maps, which are branched covers of the sphere with three branch points \cite{ITZYK,RobSanj}. This has an interpretation within topological A-model strings \cite{Gopak2011,dMKLN}. The links with tau  functions of integrable models are developed in \cite{AMMN2014}.  Matrix model formulations of general Hurwitz space problems are developed in \cite{NO2020}.  The present paper shows that  the large $N$ simplicity of the trace structures of $U(N)$ theories extends to the large $N$ simplicity of permutation invariant observables. This suggests that there may well be a  rich analogous geometrical story in the large $N$ expansion of permutation invariant models. $U(N)$ invariant models are related to two-dimensional topological field theories based on lattice gauge theories constructed from  symmetric group algebras  \cite{QuivCalc,Kimura1403}. We expect analogous developments for $S_N$ invariant models involving topological field theories based on partition algebras.   The partition functions of 
 $U(N)$ matrix models display rich large $N$ phase structures which should have interesting parallels  in the $S_N$ invariant case \cite{osti1980, Wadia1980N, Skag1984, DouglasKazakov, Sundborg2000, Aharony2004, FHY2007,Dutta2008, ramgoolam2019quiver, 2021Ali}. A recent study of $S_N$ lattice gauge theory partition functions is in \cite{BK2012}. 

Many  results in $U(N),SO(N),Sp(N)$ matrix models have been developed in the physical context of gauge-string duality.  A natural question which encompasses many of the above technical directions is whether there is a gauge-string dual interpretation for the correlators of permutation invariant observables in the simplest $O(N)$ invariant model, where we have established large $N$ factorisation.  This permutation invariant sector is one which goes beyond singlets of the continuous symmetry. Non-singlet sectors have been  organised according to more general representations of the continuous symmetry and discussed in gauge-string duality  in connection with low dimensional models of stringy black hole physics \cite{KaKoKu2001,SDTD2003,Malda0503}. It would be interesting to explore the implications of the large $N$ factorisation we have described in terms of space-time duals of this form, in particular whether there is some generalization of the connection between multi-particle states in a dual background and the factorization property along the lines of \cite{BBNS}. Double scaled matrix models, which have returned to current interest (see e.g. \cite{saad2019jt, stanford2020jt, Dionysios2020, Clifford2021})  should provide interesting settings for the investigation of permutation invariant observables in models with actions invariant under continuous symmetries. 

Finally it is worth noting that $S_N$ symmetry has been considered in the context of finite deformations of quantum mechanics in \cite{tHooft:1996cxg,BanksFinite,Kornyak, tHooft2021}. The mathematics of permutation invariant matrix models  should  have interesting interfaces with these deformations.  

\vskip.5cm 

\begin{center} 
	{ \bf Acknowledgements} 
\end{center} 
SR is supported by the STFC consolidated grant ST/P000754/1 `` String Theory, Gauge Theory \& Duality” and a Visiting Professorship at the University of the Witwatersrand, funded by a Simons Foundation grant (509116)  awarded to the Mandelstam Institute for Theoretical Physics. We are pleased to acknowledge useful conversations on the subject of this paper with Manuel Accettulli Huber, Matthew Buican, Adriana Correa, Robert de Mello Koch, Franck Gabriel, Antal Jevicki,  Rajath Radhakrishnan, Mehrnoosh Sadrzadeh and Lewis Sword.

\appendix
\section{ Parameter limits where $S_N$ invariant Gaussian models have enhanced $O(N)$ symmetry}
\label{apx: Param Limit Sage Code}
In this appendix we briefly explain the idea behind the Sage code used to find parameter limits in section \ref{sec: PIGMM}. The code can be found together with the arXiv version of the paper.

The permutation invariant Gaussian 1-matrix model defines a 11-parameter second moment
\begin{equation}
	\expval{M_{ij}M_{kl}}_{\text{PIGMM}},
\end{equation}
and the $O(N)$ model defines a 4-parameter second moment
\begin{equation}
	\expval{M_{ij}M_{kl}}_{\text{O(N)}}.
\end{equation}
The solution to the following set of $N^4$ linear equations,
\begin{equation}
	\expval{M_{ij}M_{kl}}_{\text{PIGMM}} = \expval{M_{ij}M_{kl}}_{\text{O(N)}},
\end{equation}
gives the parameter limit in the PIGMM which reconstructs the $O(N)$ model.

The above set of equations are linearly dependent. The maximal number of linearly independent equations is $11$ (the number of parameters in the PIGMM). Such a set can be constructed as follows. There are $11$ independent choices of $i,j,k,l$ in the second moment for the PIGMM, one for every inequivalent choice of setting $i,j,k,l$ equal or unequal. Two choices are equivalent if they are related by a swap $i \leftrightarrow k, j \leftrightarrow l$, or relabeling $i,j,k,l \mapsto \sigma(i), \sigma(j), \sigma(k), \sigma(l)$ for $\sigma \in S_N$.
The Sage code uses the following values for $i,j,k,l$
\begin{equation}
	\begin{array}{c c c c}
		i & j &k &l  \\ \hline
		1 & 1 & 1 &1 \\
		1 & 2 & 2 &2 \\
		2 & 1 & 2 &2 \\
		1 & 2 & 1 &2\\
		1 & 1 & 2 &2 \\
		1 & 2 & 2 &1 \\
		1 & 2 & 3 &4 \\
		1 & 2 & 1 &3 \\
		1 & 2 & 3 &2 \\
		1 & 2 & 2 &3 \\
		1 & 2 & 3 &3 
	\end{array}
\end{equation}
We repeat the same procedure for the simplest $O(N)$ model to find the parameter limit specified in equation \eqref{eqn:1in13}.
\newpage
\printbibliography
\end{document}